%% file: paper.tex
\documentclass{JHEP3}
\usepackage{hep}
\usepackage{epsfig,multicol}
\usepackage{subfigure}
\newcommand{\bs}[1]{\boldsymbol{#1}}
\def\d{\mathrm{d}}

\title{Central Exclusive Production in QCD}
\author{T.D. Coughlin\\Department of Physics \& Astronomy, University College London, Gower Street, London WC1E 6BT, UK.}
\author{J.R. Forshaw\\School of Physics \& Astronomy, University of Manchester, Manchester M13 9PL, UK.}
\preprint{MAN/HEP/2009/46}

\abstract{We investigate the theoretical description of the central
  exclusive production process, $h_1 + h_2 \to h_1+X+h_2$. Taking
  Higgs production as an example, we sum logarithmically enhanced
  corrections appearing in the perturbation series to all orders in
  the strong coupling. Our results agree with those originally
  presented by Khoze, Martin and Ryskin except that
the scale appearing in the Sudakov factor, $\mu=0.62 \sqrt{\hat{s}}$,
should be replaced with $\mu=\sqrt{\hat{s}}$, where $\sqrt{\hat{s}}$
is the invariant mass of the centrally produced system. We confirm
this result using a fixed-order calculation and show that the
replacement leads to approximately a factor 2 suppression in the
cross-section for central system masses in the range
100--500~GeV.}

\begin{document}

\input{Sections/Introduction}
\input{Sections/Durham}

\input{Sections/LowestOrder}

\input{Sections/AllOrders}

\input{Sections/ExplicitNLO}
\input{Sections/Recalculation}
\input{Sections/Impact}
\input{Sections/Conclusions}

\appendix
\input{Sections/Appendix}

\bibliographystyle{JHEP}
\bibliography{paper}

\end{document}

%% file: Sections/Introduction.tex
\section{Introduction}
%%%%%%%%%%%%%%%%%%%%%%%%%

At hadron colliders, in events producing high transverse momentum particles in the central rapidity region, the colliding particles usually break up. However, in a small fraction of events the colliding hadrons remain intact and scatter through small angles. This type of production is known as central exclusive production (CEP):
\begin{align}
	h_1(p_1) +h_2(p_2) &\to h_1(p_1') \oplus X \oplus h_2(p_2')~,
\end{align}	
where the $\oplus$ denote rapidity gaps between the outgoing hadrons and the central system $X$.

The outgoing hadron momenta may be parametrised in terms of the momentum fractions each transfers to $X$, $x_i$, and their transverse momenta, $p_{i\perp}'$:
\begin{align}
	p_1'^{\mu} &= (1-x_1) p_1^{\mu} +\frac{\boldsymbol{p}_{1\perp}'^2}{(1-x_1) s}p_2^{\mu} + p_{1\perp}'^{\mu}  \\
	p_2'^{\mu} &= (1-x_ 2) p_2^{\mu} +\frac{\boldsymbol{p}_{2\perp}'^2}{(1-x_2) s}p_1^{\mu} + p_{2\perp}'^{\mu}
\end{align}
with $s$ denoting the centre-of-mass energy squared. The CEP kinematics are defined as
\begin{align}
	\frac{\boldsymbol{p}_{i\perp}'^2}{x_i s} &\ll x_i \ll 1 \;. \label{eq:CEPKine}
\end{align}
If the outgoing hadron momenta are measured, by adding detectors far down the beam-pipe, it is possible to reconstruct the four-momentum of the central system $X$. In the CEP kinematic regime, (\ref{eq:CEPKine}), the central system's rapidity, $y$, and invariant mass squared, $\hat{s}$, are given approximately by
\begin{align}
	\hat{s} &\approx x_1x_2 s \;, \\
	y &\approx \frac{1}{2}\ln\left(  \frac{x_1}{x_2} \right) \;.
\end{align}
In addition, the process possesses a $P$-even, $C$-even, $J_z=0$ selection rule~\cite{Khoze:2000jm} (the origin of which we shall discuss subsequently). Thus CEP offers a method to measure both the mass of $X$~\cite{Albrow:2000na} (with a resolution of $\sim 2\textrm{~GeV}$ per event~\cite{Albrow:2008pn}) and its spin-parity properties~\cite{Kaidalov:2003fw}. Photon pairs~\cite{:2007na}, di-jets~\cite{Aaltonen:2007hs} and $\chi_c$ particles~\cite{Aaltonen:2009kg} produced via the CEP mechanism have now been observed at the Tevatron and there are groups within both the ATLAS and CMS collaborations actively seeking to observe these events at the LHC~\cite{Albrow:2008pn}.

For a massive enough central system, the calculation of this type of process may be performed in perturbative QCD. At lowest order in the strong coupling, two gluons in a colour singlet state are exchanged between the hadrons. Crucial however to a correct theoretical description of the process is the inclusion of logarithmically enhanced terms, appearing at all orders in the QCD perturbation series. 

In this paper we investigate the theoretical description, in perturbative QCD, of the CEP process, taking Higgs production as an example. We begin in section~\ref{sec:Durham} by introducing central exclusive production and describing the calculation by the Durham group. In section~\ref{sec:CEPLO} we discuss central exclusive Higgs production at lowest order, highlighting some of the general features of the amplitude. Then, in section~\ref{sec:CEPAllOrders}, we sum large logarithms appearing at all orders in the strong coupling. Following this analysis, we see that the Durham result requires modification and in sections~\ref{sec:FullNLO} and \ref{sec:NLO2} we describe two fixed-order calculations that confirm this result. In section~\ref{sec:Pheno}, we assess the phenomenological impact of our results. Finally, in section~\ref{sec:conclusions}, we conclude.

%% file: Sections/Durham.tex
\section{The Durham model} \label{sec:Durham}
%%%%%%%%%%%%%%%%%%%%%%%%%%%%%%%%%%%%%%%%%%%%%%%%%%%%%%%%%%%%
\begin{figure}[bt]
	\center
	\includegraphics[width=0.5\textwidth]{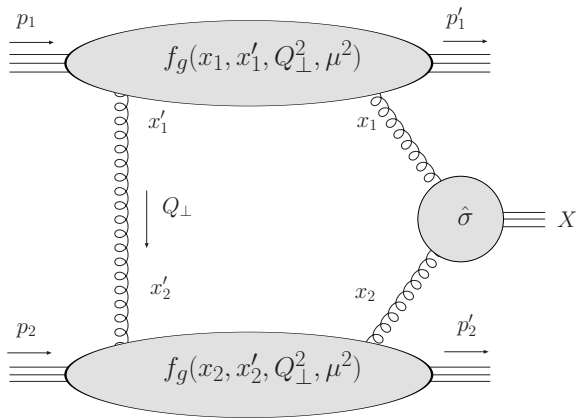}
	\caption{Schematic form of the CEP amplitude.}\label{fig:CEPschematic}
\end{figure}

The calculation of this process by the Durham group is represented schematically in figure~\ref{fig:CEPschematic}. The protons exchange a two gluon system, which must be in a colour singlet state in order that the protons remain intact. Two of the gluons then fuse to produce the central system, $X$. The cross-section is assumed to factorise in the following way~\cite{Khoze:2000cy,Khoze:2001xm}:
\begin{align}
	\frac{\partial \sigma}{\partial \hat{s} \, \partial y \, \partial \boldsymbol{p}_{1\perp}'^2 \partial \boldsymbol{p}_{2\perp}'^2} &= S^2 e^{-b(\boldsymbol{p}_{1\perp}'^2+\boldsymbol{p}_{2\perp}'^2)} \frac{\partial \mathcal{L}}{\partial \hat{s} \, \partial y} \d \hat{\sigma}(gg\to X) \;.\label{eq:Durhamdsig}
\end{align}
The transverse momenta of the final-state protons, $\boldsymbol{p}_{i\perp}'$, are assumed to be distributed according to a Gaussian, with the slope parameter, $b=4 \; \textrm{GeV}^{-2}$~\cite{Khoze:2001xm}. The factor $S^2$, known as the soft survival factor, accounts for the suppression of the cross-section due to the requirement that soft interactions between the incoming protons do not spoil the exclusivety of the process~\cite{Bjorken:1992er}. In general $S^2$ is expected to depend on the kinematics of the process, however it is common practice to set it to a constant value, corresponding to the average over the forward detector acceptance of the final-state proton transverse momenta. In any case, the factorization of $S^2$ is not expected within QCD, although there are good reasons to suppose that it holds to a reasonable approximation. We shall not consider the matter of soft survival any further in this paper and defer instead to the recent literature \cite{Bartels:2006ea,Khoze:2006uj,Strikman:2008er,Gotsman:2008tr,Gotsman:2009bn,Ryskin:2009tk}.

The partonic cross-section, $\hat{\sigma}$, is related to the matrix element for two on-shell gluons to produce the central system as
\begin{align}
	\d\hat{\sigma}(gg\to X) &= \frac{1}{2\hat{s}} \left|  \overline{\mathcal{M}}(gg\to X)\right|^2 \d\textrm{PS}_X \label{eq:CEPsigmahat}
\end{align}
where $\d\textrm{PS}_X$ is the phase-space of the final state, $X$, and
\begin{align}
	\overline{\mathcal{M}}(gg\to X) &= \frac{1}{2}\frac{1}{N^2-1} \sum_{a_1a_2}\sum_{\lambda_1\lambda_2} \delta_{a_1a_2}\delta_{\lambda_1\lambda_2} \mathcal{M}_{\lambda_1\lambda_2}^{a_1a_2}(gg\to X) \label{eq:Mbar}
\end{align}
with $\mathcal{M}_{\lambda_1\lambda_2}^{a_1a_2}$ the amplitude for two on-shell gluons, with colours $a_i$ and helicities $\lambda_i$, to fuse to produce $X$. Note that the averages are carried out at the amplitude level, in contrast to an inclusive partonic cross-section. In addition, the fact that the gluons have equal helicities ($\lambda_1=\lambda_2$) is the origin of the $J_z=0$ selection rule. 

Lastly, the effective luminosity is given by
\begin{align}
	\frac{\partial \mathcal{L}}{\partial \hat{s}\, \partial y} &= \frac{1}{\hat{s}} \left( \frac{\pi}{N^2-1}\int \! \frac{\d\boldsymbol{Q}_\perp^2}{\boldsymbol{Q}_\perp^4} f_g(x_1,x_1',\boldsymbol{Q}_\perp^2,\mu^2) f_g(x_2,x_2',\boldsymbol{Q}_\perp^2,\mu^2)  \right)^2 \;.
\end{align}
The $f_g$ are skewed, unintegrated, gluon distribution functions. Due to the kinematics of the process the amplitude is dominated by the region $x_i' \ll x_i$ and in this regime these distributions may be related to the conventional, integrated, gluon density~\cite{Martin:2001ms,Khoze:2000cy}:
\begin{align}
	f_g(x,x',\boldsymbol{Q}_\perp^2,\mu^2) \approx R_g \frac{\partial}{\partial \ln \boldsymbol{Q}_\perp^2} \left(  \sqrt{T(\boldsymbol{Q}_\perp,\mu)} x g(x,\boldsymbol{Q}_\perp^2) \right) \;.
\end{align}
The factor $R_g$  is given by
\begin{align}
	R_g &= \frac{H_g(\frac{x}{2},\frac{x}{2};\bs{Q}_\perp^2)}{x g(x,\bs{Q}_\perp^2)} \label{eq:RgDeff}
\end{align}
and accounts for the skewed effect ($H_g$ is the skewed gluon distribution, see for example~\cite{Belitsky:2005qn}). $R_g$ is approximately equal to $1.2(1.4)$ at the LHC(Tevatron)\footnote{For a LHC running at 14~TeV.}~\cite{Shuvaev:1999ce,Khoze:2001xm}. The $f_g$ distributions also include a Sudakov factor~\cite{Martin:2001ms,Dokshitzer:1978hw}:
\begin{align}
	T(\boldsymbol{Q}_\perp,\mu) &= \textrm{exp}\left(  -\int_{\boldsymbol{Q}_\perp^2}^{\hat{s}/4} \! \frac{\d k_\perp^2}{k_\perp^2}\frac{\alpha_s(k_\perp^2)}{2\pi} \int_0^{1-\Delta} \! \d z \; \left[  z P_{gg}(z) +\sum_q P_{qg}(z)  \right]  \right) \label{eq:DurhamSudakov}
\end{align}
where
\begin{align}
	\Delta &= \frac{k_\perp}{k_\perp +\mu} \label{eq:DurhamDelta} \;, \\
	\mu &= 0.62 \sqrt{\hat{s}} \;. \label{eq:DurhamMU} 
\end{align}
The Sudakov factor resums logarithmically enhanced soft and collinear virtual corrections and accounts for the fact that real radiation from the process is forbidden. 

The claim is that this expression resums logarithms in $\hat{s}/\boldsymbol{Q}_\perp^2$, to next-to-leading logarithmic accuracy. That is, it takes into account all terms of order $\alpha_s^n \ln^m(\hat{s}/\boldsymbol{Q}_\perp^2)$, with $m=2n,2n-1$. This requires a precise specification of both the lower limit on the $k_\perp$ integral in equation~(\ref{eq:DurhamSudakov}) and the cutoff on the $z$ integral as $z\to 1$. Note that the upper cutoff on the $k_\perp$ integral corresponds to non-collinear hard radiation and as such there is no logarithm associated with this region. Thus, to next-to-leading logarithmic accuracy, only its order of magnitude is required. 

The lower cutoff on the $k_\perp^2$ integral must be of the order of $\bs{Q}_\perp^2$, since radiation of a much lower transverse momentum would not be able to resolve the exchanged colour singlet system, the size of which is of order $1/|\bs{Q}_\perp|$. To extract the precise value, the Durham group use the fact that this region, with $k_\perp \sim |\bs{Q}_\perp|$ and the momentum fraction integral producing a logarithm, may be described within the BFKL framework~\cite{Fadin:1975cb,Kuraev:1976ge,Kuraev:1977fs,Balitsky:1978ic} (see for example~\cite{Forshaw:1997dc}). The BFKL summation of the momentum fraction logarithms implies the following replacement \cite{Khoze:2008cx}: 
\begin{align}
	\int \! \frac{\d ^2 k_\perp}{k_\perp^2} &\to \int \! \frac{\d^2 k_\perp}{k_\perp^2} \left(  1 - \frac{\bs{Q}_\perp^2}{k_\perp^2+(\bs{Q}_\perp-\bs{k}_\perp)^2} \right) \approx \int_{\bs{Q}^2_\perp} \frac{\d^2 k_\perp}{k^2_\perp} \label{eq:BFKLTrick}
\end{align}
Thus determining the lower limit. We shall discuss this point in more detail in section~\ref{sec:HardCollinear}. 

Having specified the lower limit in this way, the Durham group fix the cutoff on the $z$ integral, which they state is due to wide angle soft gluon radiation~\cite{Kaidalov:2003ys}, by considering the cross-section for two on-shell gluons to fuse to produce a Higgs plus one additional real gluon, which they then argue, thanks to unitarity, may be used to imply the form of the virtual corrections making up the CEP Sudakov factor. To be more specific, they consider the region $\bs{Q}_\perp^2 \ll k_\perp^2 \ll m_H^2$, leaving the $k_\perp$ integral of the radiated gluon unevaluated but integrating numerically over its polar angle. The result is then fit to a function of the form $a+b \ln(m_H/(2k_\perp))$, where $a$ and $b$ are $k_\perp$-independent constants. This approach gives the following result for the cross-section~\cite{Kaidalov:2003ys}
\begin{align}
	\sigma(gg \to Hg) & \propto \int \! \frac{\d k_\perp^2}{k_\perp^2} \frac{C_A \alpha_s}{\pi} \left( 0.212 + \ln\left(\frac{m_H}{2 k_\perp}\right) - \frac{11}{12}  \right)  \;. \label{eq:Durham0.212}
\end{align}
The factor of $11/12$ here is the usual component of the $\beta$-function coming from the $z\to 1$ finite pieces of the $P_{gg}(z)$ splitting kernel. It is stated that the factor of 0.212 is due to the region of wide angle soft gluon emission. If this is the case, it may be absorbed into the logarithmic term (which is also due to soft emission). Equivalently, it may be written in terms of a momentum fraction integral
\begin{align}
	\sigma(gg\to Hg) & \propto \int \frac{\d k_\perp^2}{k_\perp^2} \frac{\alpha_s}{2\pi} \int_0^{1-\Delta} z P_{gg}(z)~\d z
\end{align} 
where $\Delta$ is given by equations~(\ref{eq:DurhamDelta}) and (\ref{eq:DurhamMU}). Unitarity then guarantees that the soft cut-off in the real and virtual corrections must be identical, allowing one to infer the form of the $z$-integral cut-off in the Sudakov factor, (\ref{eq:DurhamSudakov}).

Our finding is that this result is not correct. Specifically, rather than equations~(\ref{eq:DurhamDelta}) and (\ref{eq:DurhamMU}) we find instead $\Delta = k_\perp/m_H$. In the next sections we will describe our evidence for this assertion: firstly in the form of an all-orders approximation to the CEP amplitude; secondly in an explicit next-to-leading order calculation of the relevant virtual corrections and thirdly, via a calculation of the $gg \to gH$ matrix element.

%% file: Sections/LowestOrder.tex
\section{Lowest order Higgs production}\label{sec:CEPLO}
%%%%%%%%%%%%%%%%%%%%%%%%%%%%%%%%%%%%%%%%%%%%%%%%%%%%%%%%%%%%%

\begin{figure}[tb]
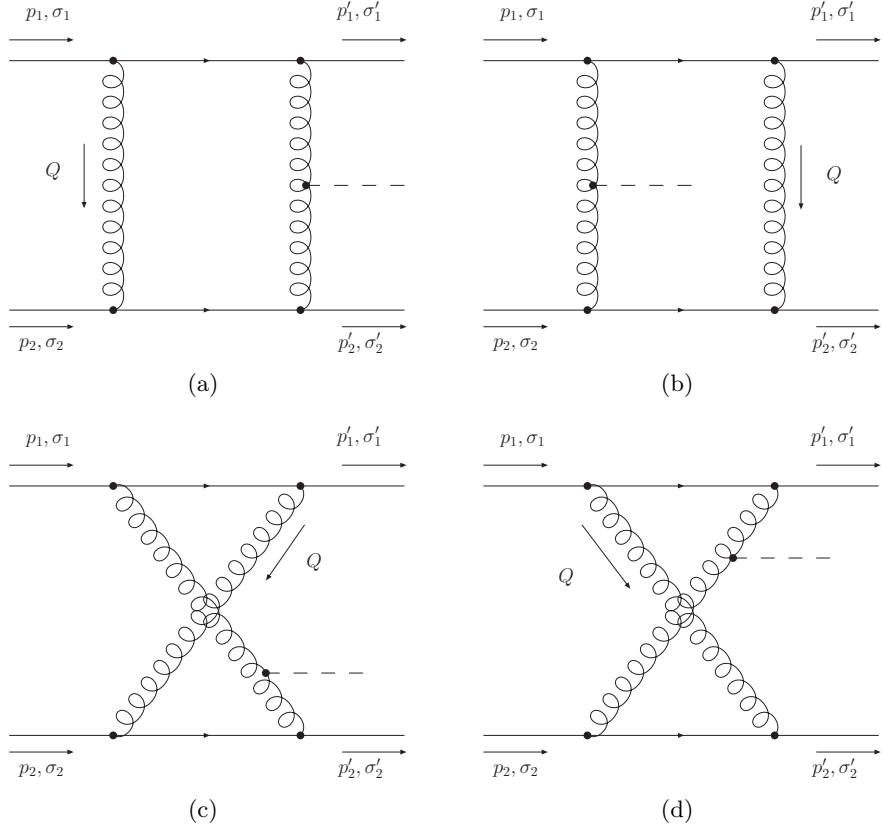

\center
\subfigure[]{\label{fig:LOa}\includegraphics[width=0.35\textwidth]{FeynmanDiagrams/CEP/LowestOrderLabeled/a/a.epsi}} \hspace{0.05\textwidth}
\subfigure[]{\label{fig:LOb}\includegraphics[width=0.35\textwidth]{FeynmanDiagrams/CEP/LowestOrderLabeled/b/b.epsi}} \\
\subfigure[]{\label{fig:LOc}\includegraphics[width=0.35\textwidth]{FeynmanDiagrams/CEP/LowestOrderLabeled/c/c.epsi}}  \hspace{0.05\textwidth}
\subfigure[]{\label{fig:LOd}\includegraphics[width=0.35\textwidth]{FeynmanDiagrams/CEP/LowestOrderLabeled/d/d.epsi}}
\caption{The lowest order diagrams contributing to $q+q'\to q\oplus H \oplus q'$, in the high energy limit.}\label{fig:LO}
\end{figure}

We begin our investigation of the Durham model by computing (in Feynman gauge) the lowest order amplitude for two quarks of different flavour to scatter into two quarks and a Higgs, $A_{\textrm{LO}}$. We take the quarks as scattering in the forward direction:
\begin{align}
	p_1'^{\mu} &= (1-x_1) p_1^{\mu} \;,\\
	p_2'^{\mu} &= (1-x_2) p_2^{\mu} \;.
\end{align}
The dependence of the amplitude on the outgoing hadron transverse momenta in equation~(\ref{eq:Durhamdsig}) is a non-perturbative effect and so we do not expect to be sensitive to it here. We will also make frequent use of the high energy limit: keeping only terms not suppressed by an inverse power of the centre-of-mass energy. In this limit, there are four lowest order diagrams which contribute, as shown in figure~\ref{fig:LO}. In addition, we work in the effective theory in which the top quark has been integrated out~\cite{Shifman:1979eb,Voloshin:1985tc,Ellis:1975ap}. This generates a point-like coupling of the Higgs to gluons (see appendix~\ref{app:largetop}).
%%%%%%%%%%%%%%%%%%%%%%%%%%%%%%%%%%
%%%%%%%%%%%%%%%%%%%%%%%%%%%%%%%%%%%%%%%%%%%%%%%

For small $x_i$ the amplitude is dominated by the region in which the exchanged gluons are soft and we may therefore take the gluons as coupling to the quark lines via eikonal vertices. Thus, for example, the contribution to the amplitude of graph~\ref{fig:LOa}, with the colour singlet contribution projected out, takes the form
\begin{align}
	A^{\ref{fig:LOa}} &= - i \delta_{\sigma_1\sigma_1'} \delta_{\sigma_2\sigma_2'}16 T_F^2 \frac{N^2-1}{N^2} g^4 p_1 \cdot p_2 \int \! \frac{\d^4Q}{(2\pi)^4} \frac{p_1^{\mu}p_2^{\nu} H_{\mu\nu}(k_1,k_2;\mu)}{\mathcal{D}}~, \label{eq:amp1a}
\end{align}
where
\begin{align}
	\mathcal{D}^{-1} &= [Q^2+i\varepsilon][(Q-p_1)^2+i\varepsilon][(Q+p_2)^2+i\varepsilon] \nonumber \\
			& \quad \times[(Q-x_1p_1)^2+i\varepsilon][(Q+x_2p_2)^2+i\varepsilon]
\end{align}
and the Higgs vertex factor, $H^{\mu\nu}$, has the form (see appendix~\ref{app:largetop})
%and the kinematics are shown in figure~\ref{fig:LOa-labeled}. The Higgs vertex factor, $H^{\mu\nu}$, has the form:
\begin{equation}
	H^{\mu\nu}(k_1,k_2;\mu) = -iC^R_1(\mu)(k_1\cdot k_2 \; g^{\mu\nu} - k_2^{\mu}k_1^{\nu}) \;.
\end{equation}
Working in the centre-of-mass frame of $p_1$ and $p_2$, with $p_1$ defining the $z$-axis, we may evaluate the $Q^+$ and $Q^-$ integrals for each graph. To leading power in the high-energy limit we obtain
\begin{align}
	A^{\ref{fig:LOa}} &= A^{\ref{fig:LOc}} = A^{\ref{fig:LOd}} = 0 \nonumber \\
	A_{\textrm{LO}} &= A^{\ref{fig:LOb}} = A_0(\mu) \int_{\Lambda} \! \frac{\d\boldsymbol{Q}^2_\perp}{\boldsymbol{Q}_\perp^4}~, \label{eq:amps-abcd}
\end{align} 
where $\Lambda$ is a cutoff imposed to define the integral and
\begin{align}
	A_0(\mu) &=  - \delta_{\sigma_1\sigma_1'}\delta_{\sigma_2\sigma_2'}C^R_1(\mu) g^4 \frac{s}{\pi} \frac{T_F^2 C_F}{N} \;.
\end{align}

Some comments are in order at this point. The first is that the amplitude is infrared power-divergent as $\Lambda \to 0$; this is to be expected, since we are dealing with a fixed-order expansion of a parton-level matrix element. As we shall see, this divergence will ultimately be tamed by a Sudakov factor (see also the footnote at the beginning of section~\ref{sec:HardCollinear}). The second comment is that the contribution to the S-matrix ($iA$) is pure imaginary. This is expected for colour singlet exchange in the high energy limit, based on arguments from Regge theory~\cite{Forshaw:1997dc}. Thirdly, note that it is perhaps not surprising that diagram \ref{fig:LOb} gives a different result from all of the other diagrams, since it is the only one in which $Q$ is pinched in the Glauber/Coulomb region~\cite{Bodwin:1981fv,Collins:1981ta}.

%%%%%%%%%%%%%%%%%%%%%%%%%%%%%%%%%%%%%%%
\begin{figure}[b]
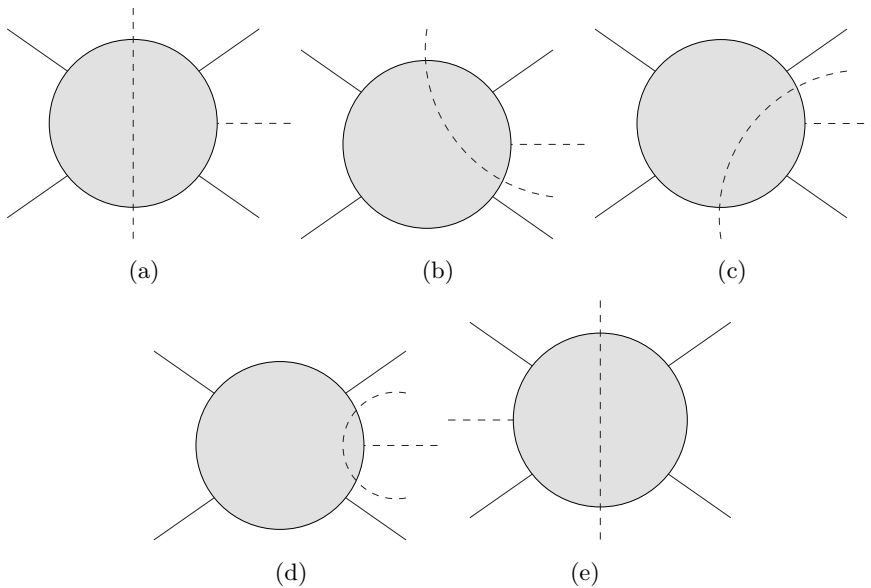

	\center
	\subfigure[]{\label{fig:CEPgeneralcuts1}\includegraphics[width=0.25\textwidth]{FeynmanDiagrams/CEPgeneralcuts/Cut1/Cut1.epsi}}
	\subfigure[]{\label{fig:CEPgeneralcuts2}\includegraphics[width=0.25\textwidth]{FeynmanDiagrams/CEPgeneralcuts/Cut2/Cut2.epsi}}
	\subfigure[]{\label{fig:CEPgeneralcuts3}\includegraphics[width=0.25\textwidth]{FeynmanDiagrams/CEPgeneralcuts/Cut3/Cut3.epsi}} \\
	\subfigure[]{\label{fig:CEPgeneralcuts4}\includegraphics[width=0.25\textwidth]{FeynmanDiagrams/CEPgeneralcuts/Cut4/Cut4.epsi}}	
	\subfigure[]{\label{fig:CEPgeneralcuts5}\includegraphics[width=0.25\textwidth]{FeynmanDiagrams/CEPgeneralcuts/Cut5/Cut5.epsi}}
	\caption{The allowed cuts of the CEP amplitude.}\label{fig:CEPgeneralcuts}
\end{figure}
%%%%%%%%%%%%%%%%%%%%%%%%%%%%%%%%%%%%%%%

Since its imaginary part dominates, we could equally well have used the Cutkosky rules~\cite{Cutkosky:1960sp} to evaluate the amplitude. In fact, it is this approach we shall use to evaluate the next-to-leading order corrections in section~\ref{sec:FullNLO}. The structure of the lowest order calculation will give us some hint as to how we may simplify the next-to-leading order corrections and so we detail it here.

\begin{figure}[t]
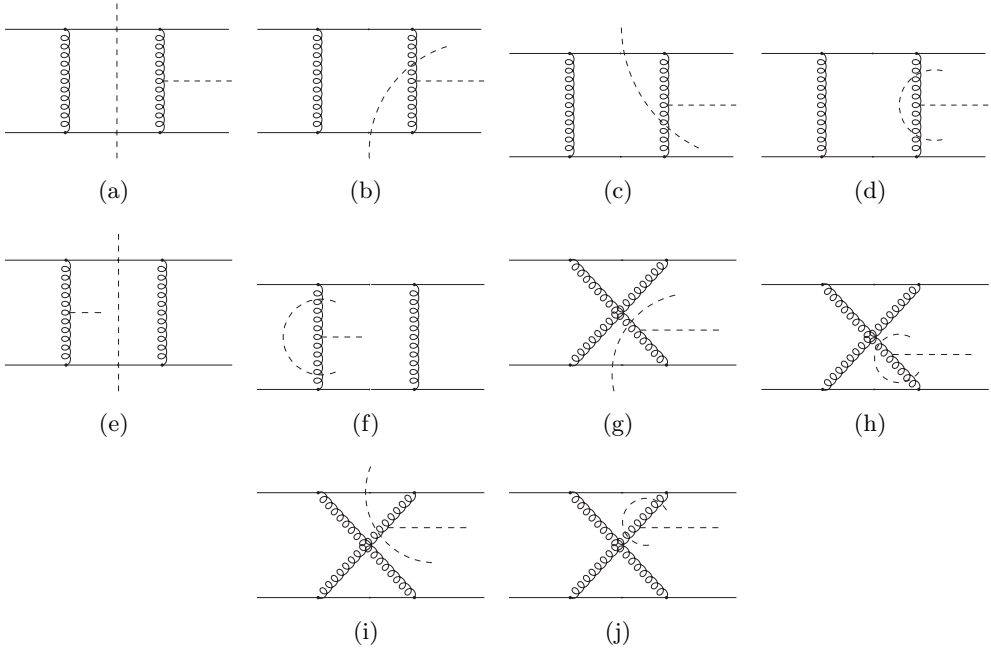

\center
\subfigure[]{\label{fig:LOcut1}\includegraphics[width=0.2\textwidth]{FeynmanDiagrams/CEP/LowestOrder/Cuts/1.epsi}}\hspace{0.2cm}
\subfigure[]{\label{fig:LOcut2}\includegraphics[width=0.2\textwidth]{FeynmanDiagrams/CEP/LowestOrder/Cuts/2.epsi}}\hspace{0.2cm}
\subfigure[]{\label{fig:LOcut3}\includegraphics[width=0.2\textwidth]{FeynmanDiagrams/CEP/LowestOrder/Cuts/3.epsi}}\hspace{0.2cm}
\subfigure[]{\label{fig:LOcut4}\includegraphics[width=0.2\textwidth]{FeynmanDiagrams/CEP/LowestOrder/Cuts/4.epsi}} \\
\subfigure[]{\label{fig:LOcut5}\includegraphics[width=0.2\textwidth]{FeynmanDiagrams/CEP/LowestOrder/Cuts/5.epsi}}\hspace{0.2cm}
\subfigure[]{\label{fig:LOcut6}\includegraphics[width=0.2\textwidth]{FeynmanDiagrams/CEP/LowestOrder/Cuts/6.epsi}}\hspace{0.2cm}
\subfigure[]{\label{fig:LOcut7}\includegraphics[width=0.2\textwidth]{FeynmanDiagrams/CEP/LowestOrder/Cuts/7.epsi}}\hspace{0.2cm}
\subfigure[]{\label{fig:LOcut8}\includegraphics[width=0.2\textwidth]{FeynmanDiagrams/CEP/LowestOrder/Cuts/8.epsi}} \\
\subfigure[]{\label{fig:LOcut9}\includegraphics[width=0.2\textwidth]{FeynmanDiagrams/CEP/LowestOrder/Cuts/9.epsi}}\hspace{0.2cm}
\subfigure[]{\label{fig:LOcut10}\includegraphics[width=0.2\textwidth]{FeynmanDiagrams/CEP/LowestOrder/Cuts/10.epsi}}
 \caption{The possible cuts of the lowest order diagrams.}\label{fig:LOcuts}
\end{figure}

Recall that the Cutkosky rules implement the unitarity relation, which states that if we sum over the cuts of a graph, $G_C$, this sum is proportional to the imaginary part of the graph without the cut, $G$:
\begin{align}
  \sum_C G_C &= 2 \Im {\mathrm{m}} (-i G) \;.\label{eq:Cutkosky}
\end{align}
For a cut to be allowed it must satisfy two conditions: the invariant-mass of the four-momenta on each side of the cut and the total energy flowing from left to right across the cut must both be positive. Given these conditions, the possible cuts of the general CEP amplitude are shown in figure~\ref{fig:CEPgeneralcuts} and the possible cuts of the lowest order diagrams are shown in figure~\ref{fig:LOcuts}. 

%%%%%%%%%%%%%%%%%%%%%%%%%%%%%%%%%%
%%%%%%%%%%%%%%%%%%%%%%%%%%%%%%%%%%%%%%%%%%%%%%%

\begin{figure}[h]
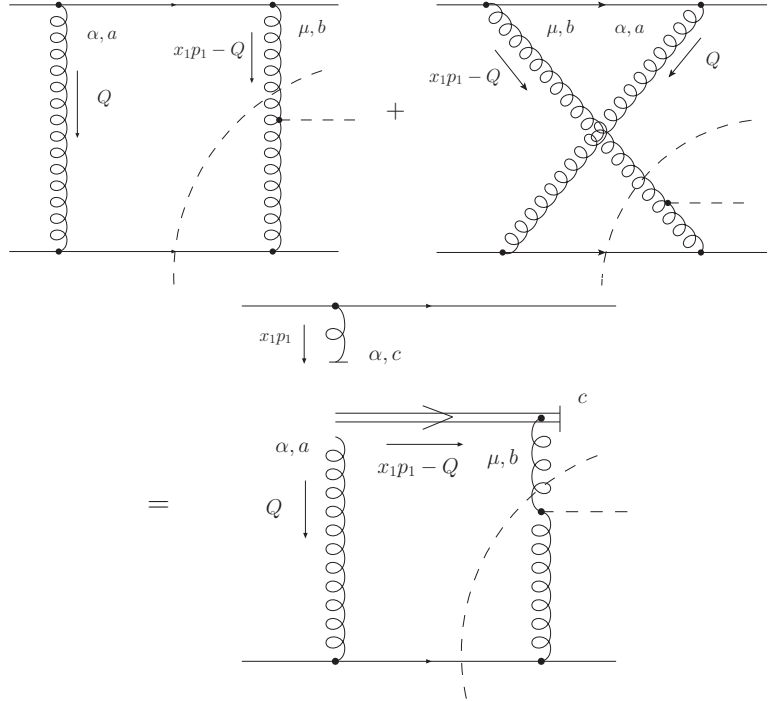

	\center
	\subfigure{\includegraphics[height=0.25\textwidth]{FeynmanDiagrams/CEPsoftapprox/1/1.epsi}} \hspace{0.005\textwidth}
	\subfigure{\includegraphics[height=0.25\textwidth]{FeynmanDiagrams/CEPsoftapprox/2/2.epsi} } \hspace{0.005\textwidth}
	\subfigure{\includegraphics[height=0.35\textwidth]{FeynmanDiagrams/CEPsoftapprox/3/3temp.epsi} }
	\caption{Rewriting a sum of cuts in terms of an insertion onto an eikonal line.}\label{fig:CEPsoftapprox}
\end{figure}

However, it turns out that we need only consider the cuts~\ref{fig:LOcut1} and \ref{fig:LOcut5}; the remaining cuts all cancel amongst one another. The reason for this is that these diagrams represent a sum over soft gluon insertions onto the upper or lower quark lines. The fact that the soft gluon in all these cases is cut ensures that it is not in the Coulomb/Glauber region and we may therefore apply the soft approximation~\cite{Collins:1985ue,Collins:1988ig,Collins:1981ta,Collins:1981uk}. The result is that we may rewrite each sum of diagrams contributing to a given cut as a single diagram with the soft gluon connected to an eikonal line. This is shown explicitly in figure~\ref{fig:CEPsoftapprox} for the diagrams~\ref{fig:LOcut2} and \ref{fig:LOcut7} (see for example~\cite{Collins:1989gx} for the eikonal Feynman rules). The two diagrams differ only in the attachments of the gluons to the upper quark lines, which give the following expressions
\begin{align}
	A^{\textrm{\ref{fig:LOcut2}}} &\propto \frac{t^bt^a}{(p_1-Q)^2}p_1^\mu p_1^\alpha \approx -\frac{t^bt^a}{2p_1 \cdot Q} p_1^\mu p_1^\alpha \\
	A^{\textrm{\ref{fig:LOcut7}}} & \propto \frac{t^at^b}{(p_1'+Q)^2}p_1^\mu p_1^\alpha \approx +\frac{t^at^b}{2p_1 \cdot Q} p_1^\mu p_1^\alpha
\end{align}
Here we took both gluons as soft, dropping all terms suppressed by $Q$ or $x_1$, which allowed us to make the eikonal approximation in the numerator and to set $p_1'\approx p_1$ in the denominator of the second diagram. In addition, the on-shell delta-function gives $|Q^2| =|2x_1p_1\cdot Q| \ll |2p_1 \cdot Q|$, allowing us to approximate the denominators as shown. The sum of diagrams is proportional to the commutator $[t^a,t^b]$ or, put another way, in the eikonal diagram of figure~\ref{fig:CEPsoftapprox} the upper quark line is connected to the lower part of the diagram by a single gluon. The key point is that we are dealing with colour singlet exchange and such diagrams contribute only to the octet exchange part of the amplitude. We may therefore neglect them.

%%%%%%%%%%%%%%%%%%%%%%%%%%%%%%%%%%%%%%%%%%%%%%
%%%%%%%%%%%%%%%%%%%%%%%%%%%%%%%%%%%%%%%%%%%%%%

This situation generalises to all orders and so we need never consider the cuts~\ref{fig:CEPgeneralcuts2}--\ref{fig:CEPgeneralcuts4} since they do not make a leading contribution to the colour singlet exchange amplitude.

%% file: Sections/AllOrders.tex
\section{All orders Higgs production} \label{sec:CEPAllOrders}
%%%%%%%%%%%%%%%%%%%%%%%%%%%%%%%%%%%%%%%%%%%%%%%%%%%%%%%%%%%%%%%
The inclusion of large logarithms appearing at all orders in the perturbation series is crucial if we are to obtain a good approximation to the CEP amplitude. Indeed, as we saw at lowest order, the result is divergent without them\footnote{Actually, the amplitude must be finite even without the inclusion of Sudakov effects, since as $\bs{Q}_\perp \to 0$ the wavelength of the exchanged gluons becomes too large to resolve the colourless protons. Such an effect requires the introduction of hadronic wavefunctions and it has been studied in~\cite{Cudell:2008gv}, using a simple non-perturbative model~\cite{Gunion:1976iy,Levin:1981rf}.}. The logarithmically enhanced terms correspond to emissions that are collinear with external particles and/or soft. In the next sub-section we shall discuss the subset of these corrections corresponding to hard collinear emissions, deferring a treatment of soft emissions to section \ref{sec:softCEPeffects}.

\subsection{Hard collinear emissions}\label{sec:HardCollinear}
%--------------------------------------------------------------------------------------

\subsubsection{Corrections that factorise into the pdfs}\label{sec:pdfs}
%.........................................................................................
 
%%%%%%%%%%%%%%%%%%%%%%%%%%%%%%%%%%%%%%%%
\begin{figure}[t]
	\center
	\includegraphics[width=0.6\textwidth]{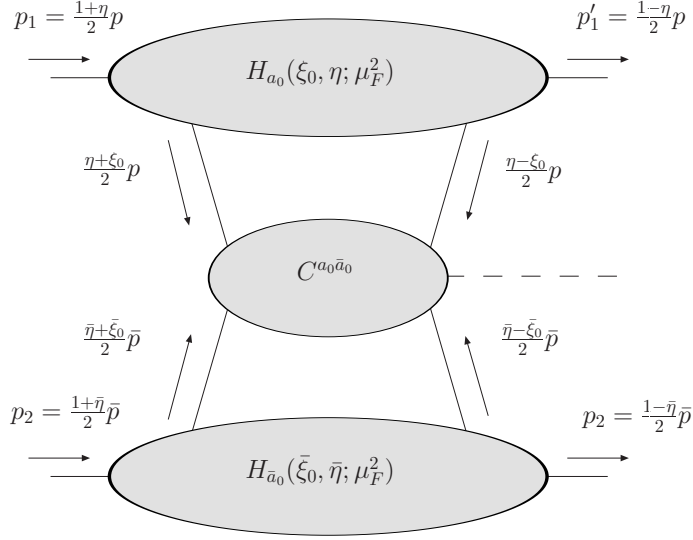}
	\caption{Factorisation of the central exclusive Higgs production amplitude.}\label{fig:Ansatz}
\end{figure}
%%%%%%%%%%%%%%%%%%%%%%%%%%%%%%%%%%%%%%%%

We begin with the ansatz that the central exclusive Higgs production amplitude, $\mathcal{A}_{\textrm{CEP}}$, factorises at some scale, $\mu_F$, much lower than all other scales in the problem. The amplitude may then be written (for small $x_1$ and $x_2$) as
\begin{align}
	\mathcal{A}_{\textrm{CEP}} &\approx 2s\sum_{a_0, \bar{a}_0}\int \! \d\xi_0 \int \! \d\bar{\xi}_0 \; H_{a_0}(\xi_0,\eta;\mu_F^2)H_{\bar{a}_0}(\bar{\xi}_0,\bar{\eta};\mu_F^2) \nonumber \\ 
	& \qquad \qquad \quad \times C^{a_0\bar{a}_0}(\xi_0,\eta,\mu_F^2;\bar{\xi}_0,\bar{\eta},\mu_F^2)~,
\end{align}
where the kinematics are shown in figure~\ref{fig:Ansatz} and we define
\begin{align}
	p^{\mu} &= p_1^{\mu}+p_1'^{\mu} \;,\\
	\bar{p}^{\mu} &= p_2^{\mu} + p_2'^{\mu} \;.
\end{align}
The large logarithmic corrections to the coefficient function, $C^{a_0\bar{a}_0}$, may be computed by cutting off the parton transverse momenta at $\mu_F$, which is equivalent to dimensional regularisation in the $\overline{\textrm{MS}}$ scheme~\cite{Catani:1990rr}. The hard collinear corrections may be organised into two transverse momentum ordered chains as follows:
\begin{align}
	C^{a_0\bar{a}_0}(\xi_0,\eta,\mu_F^2;\bar{\xi}_0,\bar{\eta},\mu_F^2) &= \sum_{n=0}^{\infty} \left[ 
\left(   -\sum_{a_n} \int_{\mu_F^2}^{\bs{Q}_\perp^2} \frac{\d l_n^2}{l_n^2} \frac{\alpha_s(l_n^2)}{4\pi}\int \d\xi_n K_{(0)}^{a_na_{n-1}}  \right)
\right. \nonumber \\
	& \qquad \cdots \left. 
\left(  -\sum_{a_1}\int_{\mu_F^2}^{l_2^2} \frac{\d l_1^2}{l_1^2}\frac{\alpha_s(l_1^2)}{4\pi} \int \d\xi_1 K_{(0)}^{a_1a_0}  \right) 
\right] \nonumber \\
	&\times \sum_{\bar{n}=0}^{\infty} \left[ 
\left(   -\sum_{\bar{a}_{\bar{n}}} \int_{\mu_F^2}^{\bar{\bs{Q}}_\perp^2} \frac{\d \bar{l}_{\bar{n}}^2}{\bar{l}_{\bar{n}}^2}  \frac{\alpha_s(\bar{l}_{\bar{n}}^2)}{4\pi}\int \d\bar{\xi}_{\bar{n}} K_{(0)}^{\bar{a}_{\bar{n}}\bar{a}_{\bar{n}-1}}  \right)
\right. \nonumber \\
	& \qquad \cdots \left. 
\left(  -\sum_{\bar{a}_1}\int_{\mu_F^2}^{\bar{l}_2^2} \frac{\d\bar{l}_1^2}{\bar{l}_1^2}\frac{\alpha_s(\bar{l}_1^2)}{4\pi} \int \d\bar{\xi}_1 K_{(0)}^{\bar{a}_1\bar{a}_0}  \right)
\right] \nonumber \\
	& \otimes C^{a_n\bar{a}_{\bar{n}}}_{\textrm{2 s-channel}}(\xi_n,\eta; \bar{\xi}_{\bar{n}},\bar{\eta}) ~,\label{eq:CollinearChain}
\end{align}
where $\bs{Q}_\perp$ and $\bar{\bs{Q}}_\perp$ are the transverse momenta of the two $s$-channel (``rung'') emissions closest to the hard scatter.  The convolution symbol $\otimes$ indicates that these momenta are integrated over (e.g. see equation~(\ref{eq:c2qq})). 
The $K^{ij}_{(0)}$ are the analogue of the DGLAP splitting functions in the case of skewed kinematics. We have omitted their arguments for clarity but it is to be understood that 
$$K^{a_i a_{i-1}}_{(0)}=K^{a_i a_{i-1}}_{(0)}\left(  \frac{\eta+\xi_{i}}{2},\frac{\eta-\xi_{i}}{2} \Big| \frac{\eta+\xi_{i-1}}{2},\frac{\eta-\xi_{i-1}}{2} \right). $$
They are given by~\cite{Belitsky:2005qn}
\begin{align}
	&K^{qq}_{(0)}(x_1,x_2|y_1,y_2) = C_F \left[  \frac{x_1}{x_1-y_1}\vartheta_{11}^0(x_1,x_1-y_1)+\frac{x_2}{x_2-y_2}\vartheta_{11}^0(x_2,x_2-y_2) \right. \nonumber \\
	& \qquad \qquad \qquad\qquad\qquad\qquad \qquad +\vartheta_{111}^0(x_1,-x_2,x_1-y_1)  \Big]_+  \;,\nonumber\\
	%K^{qg}_{(0)}(x_1,x_2|y_1,y_2) &= -\frac{n_fT_F}{2}\left[  \vartheta_{111}^1(x_1,-x_2,x_1-y_1)\left( 1+\frac{(y_1-x_1)(y_1-x_2)}{y_1y_2}\right) \right. \nonumber\\
	%& \qquad  \qquad \qquad \left. - \vartheta_{111}^0(x_1,-x_2,x_1-y_1)\frac{(x_1-y_1)}{y_1y_2}  \right] \;, \\
	&K^{qg}_{(0)}(x_1,x_2|y_1,y_2) = \frac{T_F}{2} \left[  \vartheta_{112}^1(x_1,-x_2,x_1-y_1)  +2\frac{x_1-y_1}{y_1y_2}\vartheta_{111}^0(x_1,-x_2,x_1-y_1)  \right] \nonumber \\
	&K^{gq}_{(0)}(x_1,x_2|y_1,y_2) = 2C_F \left[  (y_1-y_2)\vartheta_{111}^0(x_1,-x_2,x_1-y_1) \right. \nonumber \\
	 & \quad \qquad\qquad \qquad \qquad \qquad \qquad \qquad \left.+x_1x_2\vartheta_{111}^1(x_1,-x_2,x_1-y_1)  \right] \;, \nonumber \\
	&K^{gg}_{(0)}(x_1,x_2|y_1,y_2) = C_A \left[    \frac{x_1}{y_1}\left[  \frac{x_1}{x_1-y_1}\vartheta_{11}^0(x_1,x_1-y_1) \right]_+ \right. \nonumber \\ 
	&\qquad \quad+ \frac{x_2}{y_2}\left[  \frac{x_2}{x_2-y_2}\vartheta_{11}^0(x_2,x_2-y_2)  \right]_+  +2\frac{x_1x_2+y_1y_2}{y_1y_2}\vartheta_{111}^0(x_1,-x_2,x_1-y_1)\nonumber \\
	&\qquad \quad +2\frac{x_1x_2}{y_1y_2}\frac{x_1y_1+x_2y_2}{(x_1+x_2)^2}\vartheta_{11}^0(x_1,-x_2) +\left(\frac{1}{2}\frac{\beta_0}{C_A}+2  \right)\delta(x_1-y_1) \Big] \; , \label{eq:splittingkernels}
\end{align}
where the generalised step function is defined as
\begin{align}
	\vartheta_{\alpha_1\cdots \alpha_j}^k(x_1,\ldots ,x_j) &= \int \frac{\d\kappa}{2\pi i}\frac{\kappa^k}{\prod_{l=1}^j(x_l \kappa -1 +i \varepsilon)}
\end{align}
and the plus-distribution is given by
\begin{align}
	\left[  \frac{x_1}{x_1-y_1}\vartheta_{11}^0(x_1,x_1-y_1)  \right]_+ &= \frac{x_1}{x_1-y_1}\vartheta_{11}^0(x_1,x_1-y_1) \nonumber \\
	& \quad - \delta(x_1-y_1)\int \d x_1' \frac{x_1'}{x_1'-y_1}\vartheta_{11}^0(x_1',x_1'-y_1) \;.
\end{align}
Note that our definitions of the kernels differ slightly from those of~\cite{Belitsky:2005qn}, in particular:
\begin{align}
	K^{qq}_{(0)} &= K^{qq}_{(0)}|_{\textrm{\cite{Belitsky:2005qn}}} \qquad \qquad K^{qg}_{(0)} = \frac{1}{4 n_f}K^{qg}_{(0)}|_{\textrm{\cite{Belitsky:2005qn}}} \nonumber \\
	K^{gq}_{(0)} &=2 K^{gq}_{(0)}|_{\textrm{\cite{Belitsky:2005qn}}} \qquad \qquad K^{gg}_{(0)} =  K^{gg}_{(0)}|_{\textrm{\cite{Belitsky:2005qn}}} \;.
\end{align}

The collinear logarithms may now be absorbed into the pdfs:
\begin{align}
	\mathcal{A}_{\textrm{CEP}} &\approx 2s \sum_{a,\bar{a}} \int \d\xi \int \d\bar{\xi} \; H_a(\xi,\eta;\bs{Q}_\perp^2) H_{\bar{a}}(\bar{\xi},\bar{\eta};\bar{\bs{Q}}_\perp^2)  \; C^{a\bar{a}}_{\textrm{2 s-channel}}(\xi,\eta; \bar{\xi},\bar{\eta})
\end{align}
where
\begin{align}
	H_a(\xi,\eta;\bs{Q}_\perp^2) &= H_a(\xi,\eta;\mu_F^2) + \sum_{a_0} \int \! \d\xi_0 \; H_{a_0}(\xi_0,\eta;\mu_F^2) \nonumber \\
	& \qquad \times \sum_{n=1}^\infty \left(  -\prod_{j=1}^{n-1} \sum_{a_j} \int_{\mu_F^2}^{l_{j+1}^2}\frac{\d l_j^2}{l_j^2} \frac{\alpha_s(l_j^2)}{4\pi}\int \! \d\xi_j \; K_{(0)}^{a_ja_{j-1}}  \right) \nonumber \\
	& \qquad \times \left(   -\int_{\mu_F^2}^{\bs{Q}_\perp^2} \frac{\d l_n^2}{l_n^2} \frac{\alpha_s(l_n^2)}{4\pi} K_{(0)}^{aa_{n-1}}\left( \frac{\eta+\xi}{2},\frac{\eta-\xi}{2}\Big| \frac{\eta+\xi_{n-1}}{2},\frac{\eta-\xi_{n-1}}{2}  \right)  \right)
\end{align}
and likewise for $H_{\bar{a}}(\bar{\xi},\bar{\eta};\bar{\bs{Q}}_\perp^2)$.

%%%%%%%%%%%%%%%%%%%%%%%%%%%%%%%%%%%%%%%%
\begin{figure}[htb]
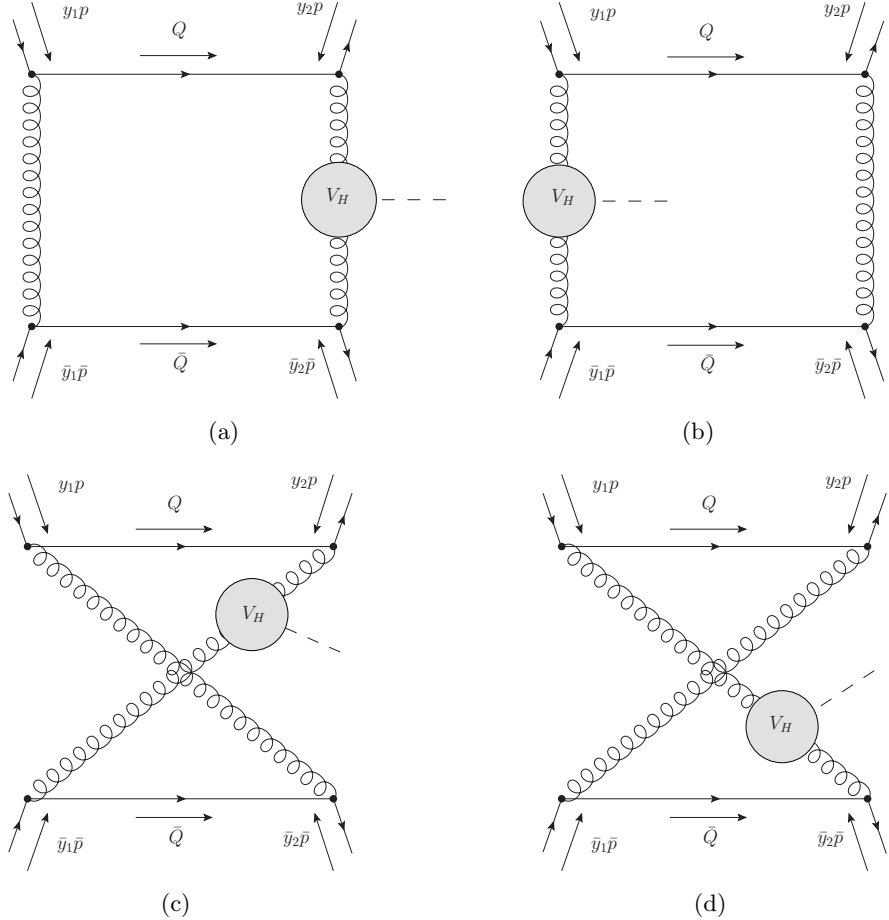

	\center
	\subfigure[]{\label{fig:Cqqa}\includegraphics[height=0.35\textwidth]{FeynmanDiagrams/HardCollinear/Cqq/1/1.epsi}} \hspace{0.05\textwidth}
	\subfigure[]{\label{fig:Cqqb}\includegraphics[height=0.35\textwidth]{FeynmanDiagrams/HardCollinear/Cqq/2/2.epsi}}  \\
	\subfigure[]{\includegraphics[height=0.35\textwidth]{FeynmanDiagrams/HardCollinear/Cqq/3/3.epsi}}  \hspace{0.15\textwidth}
	\subfigure[]{\label{fig:Cqqd}\includegraphics[height=0.35\textwidth]{FeynmanDiagrams/HardCollinear/Cqq/4/4.epsi}} 
	\caption{Diagrams contributing to $C^{qq}_{\textrm{2 s-channel}}$.}\label{fig:Cqq}
\end{figure}
%%%%%%%%%%%%%%%%%%%%%%%%%%%%%%%%%%%%%%%%

To study the coefficient function, we take $\{ a,\bar{a}\} = \{  q,q \}$ as an example. The diagrams contributing to $C^{qq}_{\textrm{2 s-channel}}$ are shown in figure~\ref{fig:Cqq}. The contribution from figure~\ref{fig:Cqqa}, for example, may be written
\begin{align}
	&C^{qq}_{\textrm{2 s-channel}}(\xi,\eta;\bar{\xi},\bar{\eta})|_{\textrm{\ref{fig:Cqqa}}} = \nonumber \\
	&\qquad \qquad \times \left(  \frac{i g^2 T_F}{4N} \sum_{a} \int \! \frac{\d^4Q}{(2\pi)^4} \frac{\textrm{Tr}\left[ \gamma^- \gamma^{\mu} \slashed{Q} \gamma^\alpha  \right]}{[Q^2+i\varepsilon][(y_1p-Q)^2+i\varepsilon][(y_2p+Q)^2+i\varepsilon] }\right) \nonumber \\
	&\qquad \qquad   \times \left( \frac{i g^2 T_F}{4N} \sum_{\bar{a}} \int \! \frac{\d^4\bar{Q}}{(2\pi)^4} \frac{\textrm{Tr}\left[ \gamma^+ \gamma^{\nu} \bar{\slashed{Q}} \gamma_\alpha  \right]}{[\bar{Q}^2+i\varepsilon][(\bar{y}_1\bar{p}-\bar{Q})^2+i\varepsilon][(\bar{y}_2\bar{p}+\bar{Q})^2+i\varepsilon]} \right) \nonumber \\
	&\qquad \qquad  \times  \frac{1}{4} i (y_1p-Q)^2 (2\pi)^4 \delta^{(4)}(y_1p+\bar{y}_1\bar{p}-Q-\bar{Q}) V_{H\mu\nu}^{a\bar{a}}(y_2p+Q,\bar{y}_2 \bar{p}+\bar{Q}) ~,
\end{align}
where $y_1 = (\eta+\xi)/2$, $y_2 =(\eta - \xi)/2$ and likewise for $\bar{y}_1,\; \bar{y}_2$. 
The factor of $1/4$ on the third line corrects for double counting after integration over $\xi$ and $\bar{\xi}$. We may approximate the delta-function as
\begin{align}
	\delta^{(4)}(y_1p+\bar{y}_1\bar{p} -Q-\bar{Q}) \approx \delta(y_1 p^+-Q^+)\delta(\bar{y}_1\bar{p}^--\bar{Q}^-) \delta^{(2)}(Q_\perp + \bar{Q}_\perp)~,
\end{align}
which decouples the momentum fraction integrals of the upper and lower sections of the diagram. In addition, if we keep only terms producing a power divergence in $Q_\perp$ and use the gauge invariance of the Higgs vertex, $V_H$, we may make the replacements
\begin{align}
	\textrm{Tr}\left[  \gamma^- \gamma^{\mu} \slashed{Q} \gamma^{\alpha}  \right] &\approx \frac{8}{p^+}(y_2-y_1)\frac{Q^\mu_\perp p^\alpha}{x_1} \;, \\
	\textrm{Tr}\left[  \gamma^+ \gamma^{\nu} \bar{\slashed{Q}} \gamma_\alpha  \right] &\approx \frac{8}{\bar{p}^-}(\bar{y}_2 -\bar{y}_1) \frac{\bar{Q}_\perp^\nu \bar{p}_\alpha}{x_2} \;.
\end{align}
Furthermore, we may write
\begin{align}
	\frac{Q_\perp^{\mu} \bar{Q}_\perp^\nu}{\bs{Q}_\perp^2} V_{H\mu\nu}^{a\bar{a}} &= \bar{V}_H \delta^{a\bar{a}} \nonumber \\
		& = \delta^{a\bar{a}} \frac{1}{2}\frac{1}{N^2-1} \sum_{a_1a_2}\sum_{\epsilon_1 \epsilon_2} \delta^{a_1a_2}\delta_{\epsilon_1 -\epsilon_2} \epsilon_1^\mu \epsilon_2^\nu V_{H\mu\nu}^{a_1a_2}~.
\end{align}
This is the origin of the requirement that the gluons fusing to produce the Higgs have equal helicities, as in equation~(\ref{eq:Mbar}).

\begin{figure}[htb]
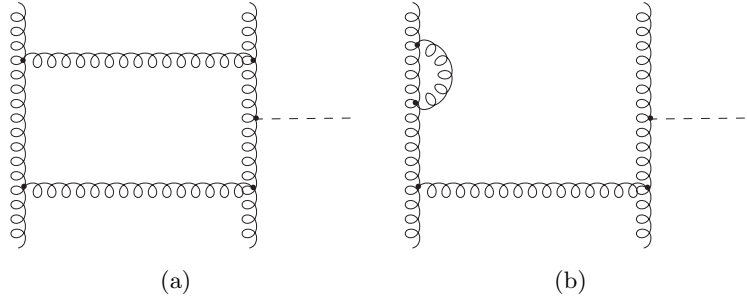

	\center
	\subfigure[]{\label{fig:PowLeading}\includegraphics[width=0.3\textwidth]{FeynmanDiagrams/HardCollinear/PowerLeading/PowerLeading.epsi}} \hspace{0.03\textwidth}
	\subfigure[]{\label{fig:PowSuppressed}\includegraphics[width=0.3\textwidth]{FeynmanDiagrams/HardCollinear/PowerSuppressed/PowerSuppressed.epsi}}
	\caption{Two diagrams contributing to CEP. The left-hand diagram makes a leading contribution, whereas the right-hand diagram is suppressed by the centre-of-mass energy.}
\end{figure}

Collecting everything together, we have
\begin{align}
	2s \; C^{qq}_{\textrm{2 s-channel}}(\xi,\eta;\bar{\xi},\bar{\eta})|_{\textrm{\ref{fig:Cqqa}}} &\approx \int \! \frac{\d \bs{Q}_\perp^2}{\bs{Q}_\perp^4} \left( -\frac{1}{2} K^{gq}\left( 0,\frac{x_1}{2}\Big| y_1,y_2  \right)  \right) \nonumber \\
	&\qquad \times \left( -\frac{1}{2} K^{gq}\left( 0,\frac{x_2}{2}\Big| \bar{y}_1,\bar{y}_2  \right)  \right)  \nonumber \\
	& \qquad \times \frac{\pi^3 2^2 (-i)}{x_1x_2(N^2-1)} \bar{V}_H \;. \label{eq:c2qq}
\end{align}
Including the other diagrams, \ref{fig:Cqqb}--\ref{fig:Cqqd} and incoming gluons, we find
\begin{align}
	\mathcal{A}_{\textrm{CEP}} &\approx \int \! \frac{\d \bs{Q}_\perp^2}{\bs{Q}_\perp^4} \sum_{a,\bar{a}} \int \d\xi \int \d\bar{\xi} \; H_a(\xi,\eta;\bs{Q}_\perp^2) H_{\bar{a}}(\bar{\xi},\bar{\eta};\bs{Q}_\perp^2) \nonumber \\
	& \times \left( -\frac{1}{2} \tilde{K}^{ga}\left( 0,\frac{x_1}{2}\Big| y_1,y_2  \right)  -\frac{1}{2} \tilde{K}^{ga}\left( \frac{x_1}{2},0\Big| y_1,y_2  \right)  \right) \nonumber \\
	& \times \left( -\frac{1}{2} \tilde{K}^{g\bar{a}}\left( 0,\frac{x_2}{2}\Big| \bar{y}_1,\bar{y}_2  \right)   -\frac{1}{2} \tilde{K}^{g\bar{a}}\left( \frac{x_2}{2},0\Big| \bar{y}_1,\bar{y}_2  \right)  \right) \nonumber \\
	& \times  \frac{\pi^3 2^2 (-i)}{x_1x_2(N^2-1)} \bar{V}_H \;. \label{eq:CEPwithKtilde}
\end{align}
Here, the $\tilde{K}$ denote the unregularised splitting kernels, i.e. the kernels of equations~(\ref{eq:splittingkernels}) but without the plus-prescription. The absence of the plus-prescription is due to a mismatch between the corrections which have the form of a self-energy and those involving the exchange of a parton in the $s$-channel. The mismatch occurs because diagrams, such as the one shown in figure~\ref{fig:PowSuppressed}, not involving at least one $s$-channel parton collinear to each hadron, are suppressed by the centre-of-mass energy. For now, we shall simply replace them with the regularised kernels but this point requires a proper treatment of soft gluon effects, which we shall cover in the next section. 

%%%%%%%%%%%%%%%%%%%%%%%%%%%%%%%%%%%%%%%%
%%%%%%%%%%%%%%%%%%%%%%%%%%%%%%%%%%%%%%%%

The pdfs satisfy an evolution equation, analagous to the DGLAP equation in non-skewed kinematics~\cite{Belitsky:2005qn}:
\begin{align}
	\frac{\partial}{\partial\ln \mu_F}\boldsymbol{H}(\xi,\eta;\mu_F^2) &= -\frac{\alpha_s}{2\pi} \int_{-1}^1 \! \d\xi' \; \boldsymbol{K}_{(0)}\left(\frac{\eta+\xi}{2},\frac{\eta-\xi}{2}\Big|\frac{\eta+\xi'}{2},\frac{\eta-\xi'}{2}\right) \boldsymbol{H}(\xi',\eta;\mu_F^2) \;, \label{eq:evolution}
\end{align}
where
\begin{align}
	\boldsymbol{H} &= \left(   \begin{array}{c} H_q \\ H_g  \end{array} \right)
\end{align}
and
\begin{align}
	\boldsymbol{K}_{(0)} &= \left(   \begin{array}{cc} K_{(0)}^{qq} & K_{(0)}^{qg} \\ K_{(0)}^{gq} & K_{(0)}^{gg} \end{array} \right) \;.
\end{align}
We may then use this equation and the symmetry relation, $H_g(-\xi,\eta;\bs{Q}_\perp^2)=H_g(\xi,\eta;\bs{Q}_\perp^2)$~\cite{Belitsky:2005qn}, to write
\begin{align}
	\mathcal{A}_{\textrm{CEP}} &\approx \int \! \frac{\d\bs{Q}_\perp^2}{\bs{Q}_\perp^4}  \frac{\partial}{\partial \ln(\bs{Q}_\perp^2)} \left[ H_g\left(\frac{x_1}{2},\frac{x_1}{2};\bs{Q}_\perp^2\right)\right]  \frac{\partial}{\partial \ln(\bs{Q}_\perp^2)} \left[ H_g\left(\frac{x_2}{2},\frac{x_2}{2};\bs{Q}_\perp^2\right) \right] \nonumber \\
	& \qquad \qquad \qquad \times \frac{\pi^3 2^4 (-i)}{x_1x_2(N^2-1)} \bar{V}_H \;. \label{eq:CEPpdfderiv}
\end{align}

\subsubsection{Corrections that generate the Sudakov factor}\label{sec:suda}
%...............................................................................................................

%%%%%%%%%%%%%%%%%%%%%%%%%%%%%%%%%%%%%%%%
\begin{figure}[htb]
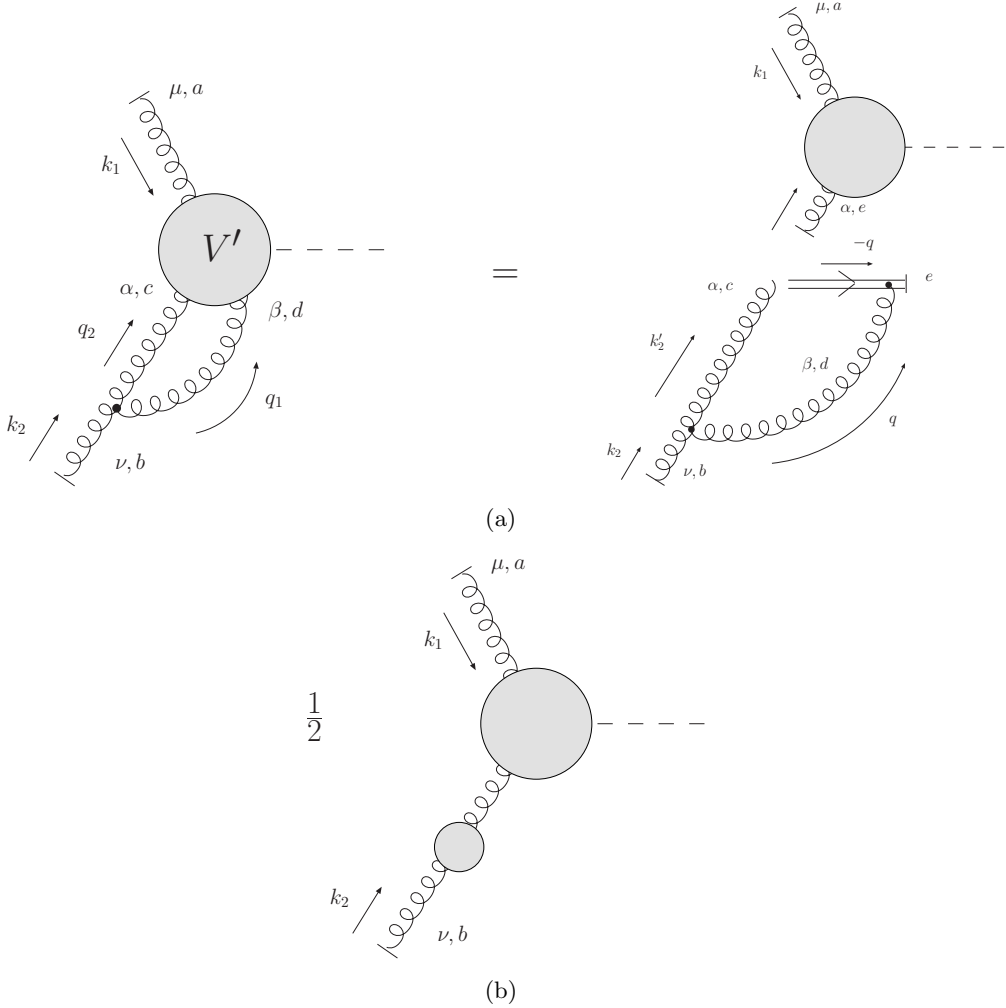

	\center
	\subfigure[]{\label{fig:Sudakova}\includegraphics[width=0.45\textwidth]{ExtraDiagrams/1/1.epsi} \hspace{0.06\textwidth}
	\includegraphics[width=0.35\textwidth]{FeynmanDiagrams/SudakovDiagrams/a/a.epsi}}   \\
	\subfigure[]{\label{fig:Sudakovb}\includegraphics[width=0.35\textwidth]{FeynmanDiagrams/SudakovDiagrams/b/b.epsi}} 
	\caption{Form of large logarithmic corrections to $V_H$ due to emissions collinear to $k_2$.}\label{fig:SudakovDiagrams}
\end{figure}
%%%%%%%%%%%%%%%%%%%%%%%%%%%%%%%%%%%%%%%%

Although we have now dealt with all emissions that factorise into the pdfs, there are still large logarithms contained in $\bar{V}_H$. Again, we deal first with the hard collinear logarithms, deferring the treatment of the soft region to section~\ref{sec:softCEPeffects}.

The diagrams contributing a collinear logarithm with respect to $k_2$ are shown in figure~\ref{fig:SudakovDiagrams}, with equivalent diagrams for $q$ collinear to $k_1$ not shown. Dealing with diagram~\ref{fig:Sudakova} first, we obtain
\begin{align}
	V^{ab}_{H\mu\nu}|_{\textrm{\ref{fig:Sudakova}}} &= \int \! \frac{\d^4q}{(2\pi)^4} \frac{(-i)}{q_1^2+i\varepsilon}\frac{(-i)}{q_2^2+i\varepsilon}f^{cbd}V_{3\alpha\nu\beta}(-q_2,k_2,-q_1) V_\mu'^{\beta\alpha;acd} \;. 
\end{align}
The three-gluon vertex is
\begin{align}
	V_{3\alpha\nu\beta}(-q_2,k_2,-q_1) &= g( g_{\alpha\nu}(-q_2-k_2)_\beta +g_{\nu\beta}(k_2+q_1)_\alpha +g_{\beta\alpha}(q_2-q_1)_\nu ) \nonumber \\
	&\to g (-(1+z_1)g_{\alpha\nu}k_{2\beta} + (1+z_2)g_{\beta\nu}k_{2\alpha})  \nonumber \\
	& \approx g\left(  -\frac{1+z_1}{1-z_1} g_{\alpha\nu} q_{1\beta} +\frac{1+z_2}{1-z_2}g_{\beta\nu}q_{2\alpha} \right)  \label{eq:zPggEikonal}
\end{align}
where we introduced the following Sudakov decomposition of the $q_i$:
\begin{align}
	q_i^{\mu} &= (1-z_i) \hat{k}_2^\mu +\beta_i v^\mu + q_{i\perp} 
\end{align}
with
\begin{align}
	\hat{k}_2^\mu &= k_2^\mu - \frac{k_2^2}{2 k_2 \cdot v} v^\mu\;,  \qquad \quad \hat{k}^2_2 =v^2=0 \;.
\end{align}
In equation~(\ref{eq:zPggEikonal}) we first used the fact that we may set $q_i^{\mu} \approx (1-z_i) k_2^\mu$ in the numerator, up to terms that do not generate a logarithm. Following this, we used the transversality of the incoming gluons to set $k_2^\nu \to 0$. We also multiplied and divided by $1-z_i$. This last step is perfectly valid for hard collinear emissions, however, it will require modification when we come to consider soft effects.

Now observe that $q_{1\beta} V_\mu'^{\beta\alpha;acd}$ and $q_{2\alpha}V_\mu'^{\beta\alpha;acd}$ describe a set of diagrams with an external gluon attached whose polarisation vector is in the direction of its momentum, i.e. it is longitudinally polarised. The QCD Ward identity may then be applied to factor this gluon from $V'$, onto an eikonal line (see for example~\cite{Collins:1989gx}). This is shown in figure~\ref{fig:Sudakova}  and gives the following expression
\begin{equation}
	V_{H\mu\nu}^{ab}|_{\textrm{\ref{fig:Sudakova}}}  = \int \! \frac{\d^4q}{(2\pi)^4} \frac{(-i)}{k_2'^2+i\varepsilon}\frac{(-i)}{q^2+i\varepsilon}f^{cbd}(-gg_{\alpha\nu}(1+z)k_{2\beta} ) \frac{i}{(-q\cdot v)} igv^{\beta}(-if^{dec})V_{H\mu\alpha}^{ae}
\end{equation}
where 
\begin{align}
	q^{\mu}&=(1-z)\hat{k}_2^\mu +\beta v^\mu +q_\perp^\mu \;.
\end{align}
Now, doing the $\beta$ integral by contour integration, we find
\begin{align}
	V_{H\mu\nu}^{ab}|_{\textrm{\ref{fig:Sudakova}}} & \approx - \frac{\alpha_s}{4\pi} C_A \int_0^1 \! \d z \int_{(1-z)z \bs{Q}_\perp^2} \frac{\d l_1^2}{l_1^2} \frac{(1+z)}{1-z} V_{H\mu\nu}^{ab}  \;, \label{eq:wrong}
\end{align}
with $l_1^2 = \bs{q}^2_\perp -(1-z)z k_2^2$ and we made the replacement $k_2^2\approx -\bs{Q}_\perp^2$. Note that, since we are only concerned with the logarithmic terms, we may replace the lower limit on the $l_1^2$ integral as $(1-z)z \bs{Q}_\perp^2 \to (1-z) \bs{Q}_\perp^2$. 

Before discussing the contribution from figure~\ref{fig:Sudakovb}, we note that these logarithmic corrections are going to generate the Sudakov factors, however, observe that the transverse momentum integral extends down to $(1-z)\bs{Q}_\perp^2$. This is in contrast to the integral in the Sudakov factor, which is cut off at $\bs{Q}_\perp^2$, though note that the difference is only relevant for the $(1-z)^{-1}$ term coming from diagram~\ref{fig:Sudakova}. It appears then that, if we cut off the $z$-integral at $1-z \sim |\bs{q}_\perp|/m_H$ (see section~\ref{sec:softCEPeffects}), this piece will generate a Sudakov factor with twice the double logarithmic contribution of equation~(\ref{eq:DurhamSudakov}). This is not the case however. 

\begin{figure}[tbh]
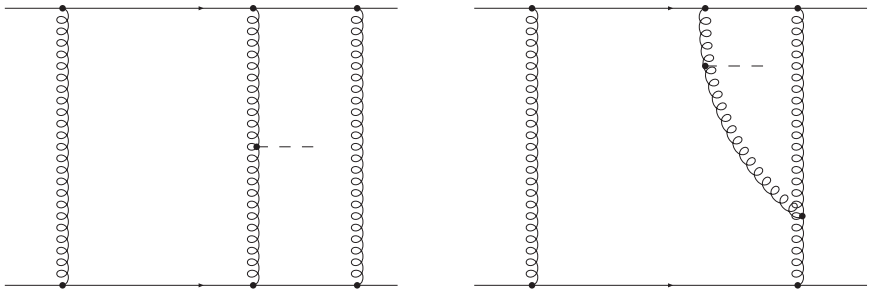

	\center
	\subfigure{\label{fig:Nonptordereda}\includegraphics[height=0.25\textwidth]{ExtraDiagrams/2/2.epsi}} \hspace{0.05\textwidth}
	\subfigure{\label{fig:Nonptorderedb}\includegraphics[height=0.25\textwidth]{ExtraDiagrams/3/3.epsi} } 
	\caption{An example of diagrams generating a logarithm not included in the transverse momentum ordered approximation.}\label{fig:Nonptordered}
\end{figure}

Working in the transverse momentum ordered approximation (see equation~(\ref{eq:CollinearChain})), we have missed a contribution from diagrams like those shown in figure~\ref{fig:Nonptordereda} and \ref{fig:Nonptorderedb}. Clearly the diagrams in figure \ref{fig:SudakovDiagrams} cannot account for the screening of long wavelength emissions which results from the colour neutrality of the $t$-channel exchange (since they are oblivious to the role of the screening gluon). Fortunately,
the contribution from diagrams like those of figure~\ref{fig:Nonptordered}, in the region $\bs{q}_\perp^2<\bs{Q}_\perp^2$, with the $(1-z)^{-1}$ piece generating a logarithm, is included in the BFKL corrections. As we shall now see, the BFKL equation guarantees that there is in fact no large logarithm generated by the transverse momentum integral in this region and fixes the lower limit on the $l_1^2$ integral in equation (\ref{eq:wrong}) at $\bs{Q}_\perp^2$ rather than $(1-z)\bs{Q}_\perp^2$.

%%%%%%%%%%%%%%%%%%%%%%%%%%%%%%%%%%%%%%%%%%%%%%
%%%%%%%%%%%%%%%%%%%%%%%%%%%%%%%%%%%%%%%%%%%%%%

%%%%%%%%%%%%%%%%%%%%%%%%%%%%%%%%%%%%%%%%%%%%%%
\begin{figure}[tbh]
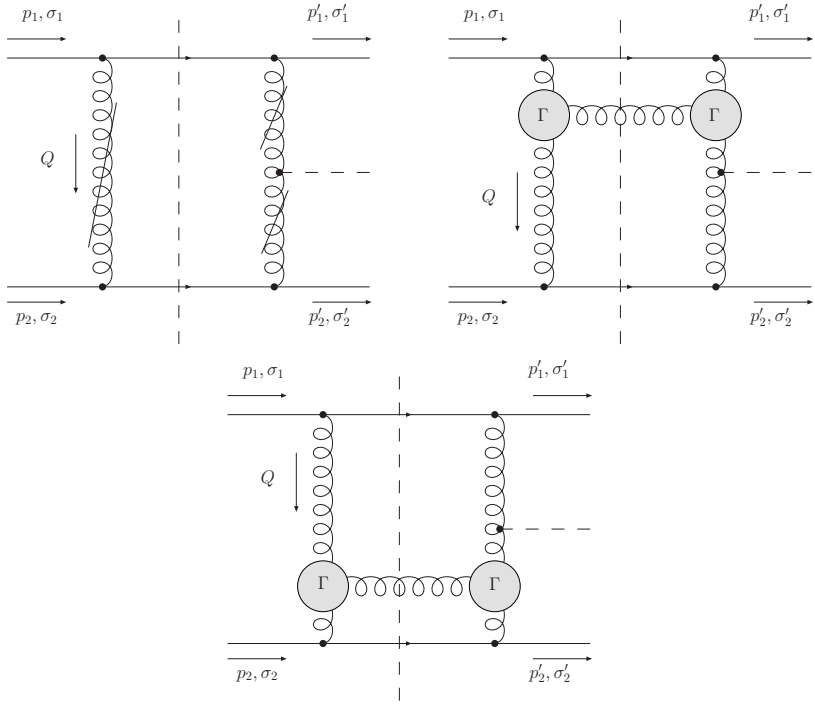

	\center
	\subfigure{\includegraphics[height=0.3\textwidth]{FeynmanDiagrams/BFKLDiagrams/1/1.epsi}} \hspace{0.05\textwidth}
	\subfigure{\includegraphics[height=0.3\textwidth]{FeynmanDiagrams/BFKLDiagrams/2/2.epsi} } \hspace{0.05\textwidth}
	\subfigure{\includegraphics[height=0.3\textwidth]{FeynmanDiagrams/BFKLDiagrams/3/3.epsi} }
	\caption{Next-to-leading order corrections to quark-quark central exclusive production in the BFKL formalism. Slashed gluon propagators and vertices labelled $\Gamma$ indicate reggeised gluons and Lipatov vertices respectively, see~\cite{Forshaw:1997dc}. Not shown are diagrams with the Higgs to the left of the cut.}\label{fig:NLOBFKL}
\end{figure}
%%%%%%%%%%%%%%%%%%%%%%%%%%%%%%%%%%%%%%%%%%%%%%

Consider the one-loop corrections to $C^{qq}$ in the BFKL region, displayed in figure~\ref{fig:NLOBFKL}, where we calculate the amplitude from cuts as described in section~\ref{sec:CEPLO}. In this region, the full set of virtual corrections to either side of the cut
% which include the $|\bs{q}_\perp| \sim |\bs{Q}_\perp|$ piece of $\bar{V}_H$ we have been discussing, 
are summed up in the reggeised gluon propagator (denoted by a slash). These corrections amount to the replacement, in the lowest order graphs, of
\begin{align}
	\frac{1}{Q^2(x_1p_1-Q)^2 (x_2p_2+Q)^2}  &\to \frac{1}{Q^2(x_1p_1-Q)^2 (x_2p_2+Q)^2} \left(\frac{s}{\bs{Q}_\perp^2}  \right)^{2\epsilon_G(Q^2)}~, \label{eq:ReggeReplacement}
\end{align}
where the gluon Regge trajectory, $\epsilon_G(Q^2)$, is given by
\begin{align}
	\epsilon_G(Q^2) &= -\frac{C_A \alpha_s}{(2\pi)^{d-2}} \int \! \d^{d-2}\bs{k}_\perp \frac{\bs{Q}_\perp^2}{\bs{k}_\perp^2(\bs{k}_\perp-\bs{Q}_\perp)^2} \nonumber \\
	& = -\frac{2C_A\alpha_s}{(2\pi)^{d-2}} \int \! \d^{d-2}\bs{k}_\perp \frac{\bs{Q}_\perp^2}{(\bs{k}_\perp-\bs{Q}_\perp)^2[\bs{k}_\perp^2+(\bs{k}_\perp-\bs{Q}_\perp)^2]} \nonumber \\
	& = \frac{C_A\alpha_s}{2\pi}\frac{(4\pi)^\epsilon}{\Gamma(1-\epsilon)} \frac{(\bs{Q}_\perp^2)^{-\epsilon}}{\epsilon} \label{eq:GluonRegge}\;.
\end{align}
%Note that the transverse momentum integral in $\epsilon_G(Q^2)$ only generates a logarithm for $\bs{k}_\perp^2<\bs{Q}_\perp^2$. 
%So as stated, these corrections only include this region of the correction to $\bar{V}_H$. 
Note that there is no double logarithm. Moreover, the region $\bs{k}_\perp^2<\bs{Q}_\perp^2$ generates only single logarithms in $s/\bs{Q}_\perp^2$, i.e. it does not lead to any logarithms in $m_H^2/\bs{Q}_\perp^2$. The complementary region,  $\bs{k}_\perp^2>\bs{Q}_\perp^2$ does not generate a transverse momentum logarithm and as such it is never able to generate a logarithm in the Higgs mass (although it does contribute to a logarithm in $s$, which can be absorbed into the pdfs). Hence we are justified in fixing the infra-red cutoff in equation (\ref{eq:wrong}) equal to $\bs{Q}_\perp^2$.  It will be instructive to observe how the elimination of the would-be double logs and the corresponding gluon reggeisation comes about in section~\ref{sec:FullNLO}, where we present an explicit calculation of the next-to-leading order corrections to $C^{qq}$. 

%%%%%%%%%%%%%%%%%%%%%%%%%%%%%%%%%%%%%%%%%%%%%%
\begin{figure}[tb]
	\center
	\includegraphics[height=0.55\textwidth]{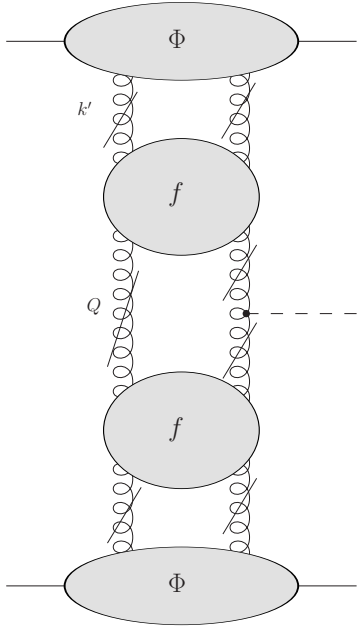}
	\caption{Schematic form of the central exclusive production amplitude in the BFKL formalism. The functions $f$ are related to the four-gluon Green function and $\Phi$ are proton impact factors, see~\cite{Forshaw:1997dc}.}\label{fig:CEPBFKL}
\end{figure}
%%%%%%%%%%%%%%%%%%%%%%%%%%%%%%%%%%%%%%%%%%%%%%

%%%%%%%%%%%%%%%%%%%%%%%%%%%%%%%%%%%%%%%%%%%%%%
\begin{figure}[b]
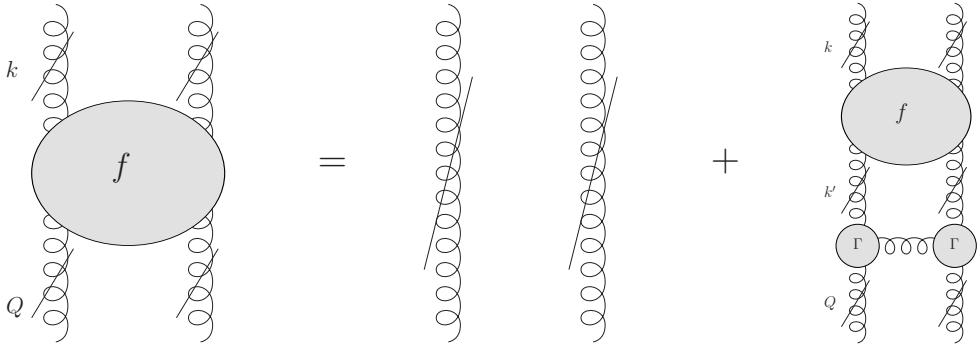

	\center
	\subfigure{\includegraphics[height=0.3\textwidth]{FeynmanDiagrams/BFKLDiagrams/5/5.epsi}} \hspace{0.05\textwidth}
	\subfigure{\includegraphics[height=0.3\textwidth]{FeynmanDiagrams/BFKLDiagrams/6/6.epsi} } \hspace{0.05\textwidth}
	\subfigure{\includegraphics[height=0.3\textwidth]{FeynmanDiagrams/BFKLDiagrams/7/7.epsi} }
	\caption{The BFKL integral equation in diagrammatic form.}\label{fig:BFKLequation}
\end{figure}
%%%%%%%%%%%%%%%%%%%%%%%%%%%%%%%%%%%%%%%%%%%%%%

We will close this discussion on BFKL by showing that the would-be infra-red divergence arising from $\epsilon_G(Q^2)$ is cancelled by emissions across the cut, which can be written in terms of the Lipatov vertex, $\Gamma_{\mu\nu}^\sigma$ (see \cite{Forshaw:1997dc} equation~(3.11)). After re-summation, the BFKL corrections can be absorbed into a function, $f(\omega,\bs{k}_\perp,\bs{Q}_\perp)$, which is related to the Green function with four off-shell gluons (see \cite{Forshaw:1997dc} equation~(4.8))\footnote{$\omega$ is a Mellin transform variable, conjugate to the centre-of-mass energy.}. The amplitude may then be computed after convolution with some non-perturbative function containing the long-distance physics associated with the external hadrons, as depicted schematically in figure~\ref{fig:CEPBFKL}.  $f(\omega,\bs{k}_\perp,\bs{Q}_\perp)$ obeys an integral equation, the BFKL equation, shown diagrammatically in figure~\ref{fig:BFKLequation}, i.e.
\begin{align}
	\omega f(\omega,\bs{Q}_\perp,\bs{k}_\perp) &= \delta^{(2)}(\bs{Q}_\perp - \bs{k}_\perp) + \frac{C_A\alpha_s}{\pi^2}\int \! \frac{\d^2\bs{k}_\perp'}{(\bs{k}_\perp'-\bs{Q}_\perp)^2} f(\omega,\bs{k}_\perp',\bs{k}_\perp)  \nonumber \\
	& \qquad +2 \epsilon_G(-\bs{Q}_\perp^2) f(\omega,\bs{Q}_\perp,\bs{k}_\perp) \nonumber \\
	&=  \delta^{(2)}(\bs{Q}_\perp - \bs{k}_\perp)  \nonumber \\
	& \qquad +\frac{C_A \alpha_s}{\pi^2} \int \! \frac{\d^2\bs{k}_\perp'}{(\bs{k}_\perp'-\bs{Q}_\perp)^2} \left(  f(\omega,\bs{k}_\perp',\bs{k}_\perp)  - \frac{\bs{Q}_\perp^2 f(\omega,\bs{Q}_\perp,\bs{k}_\perp)}{[\bs{k}_\perp'^2 + (\bs{k}_\perp'-\bs{Q}_\perp)^2]}  \right)~.
\end{align}
Now observe that, for $\bs{k}_\perp'^2 \ll \bs{Q}_\perp^2$, the first and second terms in parenthesis (the first term corresponds to emissions across the cut) cancel one another. 

The above is essentially the same argument, though presented in slightly different terms, that the Durham group use to set the lower limit on the Sudakov factor in equation~(\ref{eq:BFKLTrick}). We will verify this argument explicitly in section~\ref{sec:FullNLO}, where we detail the results of a full next-to-leading order calculation of the virtual corrections which contribute to the Sudakov factor.

Returning to the calculation of diagram~\ref{fig:Sudakova}, we find
\begin{align}
	V_{H\mu\nu}^{ab}|_{\textrm{\ref{fig:Sudakova}}} & \approx - \frac{\alpha_s}{4\pi} C_A \int_0^1 \! \d z \int_{\bs{Q}_\perp^2}^{l_2^2} \frac{\d l_1^2}{l_1^2} \frac{(1+z)}{1-z} V_{H\mu\nu}^{ab}  \;.
\end{align}
In addition to replacing the lower limit, we have now made explicit the upper limit of the $l_1^2$ integration. The variable $l_2$ is the analogue of $l_1$, but for the next emission contained in $V_H$. 

Next, the one-loop gluon propagator corrections (see for example~\cite{Grozin:2007zz}) give, for figure~\ref{fig:Sudakovb},
\begin{align}
	V_{H\mu\nu}^{ab}|_{\textrm{\ref{fig:Sudakovb}}} & \approx \frac{\alpha_s}{4\pi} \left( \frac{10}{12}C_A - \frac{2 T_F n_f}{3}  \right) \int_{-k_2^2}^{\mu_R^2} \! \frac{\d l_1^2}{l_1^2} V_{H\mu\nu}^{ab}~,
\end{align}
where $\mu_R$ is the renormalisation scale. By setting $\mu_R^2=l_2^2$, the two contributions in figure~\ref{fig:SudakovDiagrams} may be rewritten in terms of the DGLAP splitting functions, with the additional effect that the strong coupling now runs with $l_1^2$:
\begin{align}
	\bar{V}_H|_{n\textrm{ collinear emissions}} &= -2 \int^{l_2^2}_{\bs{Q}_\perp^2} \frac{\d l_1^2}{l_1^2} \frac{\alpha_s(l_1^2)}{4\pi} \int_0^1 \d z \left(  z \tilde{P}_{gg}(z) +n_f P_{qg}(z) \right) \nonumber \\
	& \qquad \qquad \qquad \qquad  \times  \bar{V}_H|_{n-1\textrm{ collinear emissions}}\label{eq:VHcollinear}
\end{align}
and we also included the contribution from emissions collinear to $k_1$. The tilde again denotes the splitting function without the plus-prescription.

Equation~(\ref{eq:VHcollinear}) is still incorrect as it stands, since the integral diverges in the soft limit, $z \to 1$. This is an artifact of our hard collinear approximation: for fixed, finite, transverse momentum, a soft gluon's energy cannot vanish. We shall discuss how this problem is rectified in the next section. For now, we ignore the divergence and iterate (\ref{eq:VHcollinear}) until we are left with the tree level vertex:
\begin{align}
	\bar{V}_H &= \sum_{n=0}^{\infty} \prod_{i=1}^n \left(   - \int_{\bs{Q}_\perp^2}^{l_{i+1}^2}  \frac{\d l_i^2}{l_i^2} \frac{\alpha_s(l_i^2)}{4\pi} \int_0^1 \d z \left( z\tilde{P}_{gg}(z) +n_f P_{qg}(z)  \right)  \right) \nonumber \\
	& \quad \times \sum_{\bar{n}=0}^{\infty} \prod_{\bar{i}=1}^{\bar{n}} \left(   - \int_{\bs{Q}_\perp^2}^{\bar{l}_{\bar{i}+1}^2} \frac{\d\bar{l}_{\bar{i}}^2}{\bar{l}_{\bar{i}}^2} \frac{\alpha_s(\bar{l}_{\bar{i}}^2)}{4\pi} \int_0^1 \d z \left( z\tilde{P}_{gg}(z) +n_f P_{qg}(z)  \right)  \right) \nonumber \\
	& \quad \times \overline{\mathcal{M}}(gg \to H)~, \label{eq:VHemissions}
\end{align}
where $l_{n+1}^2=\bar{l}_{\bar{n}+1}^2 =m_H^2$ and $\overline{\mathcal{M}}$ is as defined in equation~(\ref{eq:Mbar}). We may then use the following identity for ordered integrals
\begin{align}
	\int_{\bs{Q}_\perp^2}^{l_2^2} \d l_1^2 \cdots \int_{\bs{Q}_\perp^2}^{m_H^2} \d l_n^2 &= \frac{1}{n!}\prod_{j=1}^n \int_{\bs{Q}_\perp^2}^{m_H^2} \d l_j^2
\end{align}
which allows us to rewrite equation~(\ref{eq:VHemissions}) in terms of exponentials:
\begin{align}
	\bar{V}_H &= \exp \left[  -\int_{\bs{Q}_\perp^2}^{m_H^2} \frac{\d l^2}{l^2} \frac{\alpha_s(l^2)}{4\pi} \int_0^1 \d z \left(  z \tilde{P}_{gg}(z) +n_f P_{qg}(z)  \right) \right] \nonumber \\
	& \quad \times  \exp \left[  -\int_{\bs{Q}_\perp^2}^{m_H^2} \frac{\d \bar{l}^2}{\bar{l}^2} \frac{\alpha_s(\bar{l}^2)}{4\pi} \int_0^1 \d z \left(  z \tilde{P}_{gg}(z) +n_f P_{qg}(z)  \right) \right] \nonumber \\
	& \quad \times \overline{\mathcal{M}}(gg \to H) \;. \label{eq:divergentSudakov}
\end{align}
Thus we begin to see the emergence of the Sudakov factors.

\subsection{Soft emissions} \label{sec:softCEPeffects}
%--------------------------------
So far we have deferred the treatment of soft gluon effects: In section \ref{sec:pdfs} we noted that the final rung before the hard scatter involved an unregulated splitting function, e.g. see equation (\ref{eq:CEPwithKtilde}) and in section \ref{sec:suda} we noted a potential divergence in the $z \to 1$ limit of the Sudakov factor, e.g. see equation (\ref{eq:VHemissions}). We shall now turn our attention to these matters and consider the soft gluon limit more carefully than hitherto.

\subsubsection{The Sudakov factor}
%..............................................................

%%%%%%%%%%%%%%%%%%%%%%%%%%%%%%%%%%%
\begin{figure}[htb]
	\center
	\includegraphics[width=0.35\textwidth]{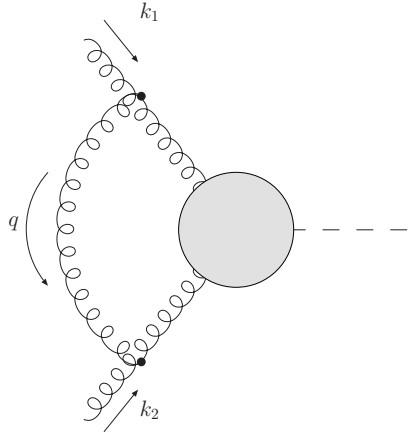}
	\caption{Form of the softest gluon attachment to the Higgs vertex.}\label{fig:softHiggsvertex}
\end{figure}
%%%%%%%%%%%%%%%%%%%%%%%%%%%%%%%%%%%%

We shall begin our discussion of soft effects by dealing with the $z \to 1$ divergence in the Higgs vertex corrections. The softest gluon must attach to both gluons fusing to produce the Higgs, as shown in figure~\ref{fig:softHiggsvertex}. This contribution has the form
\begin{align}
	\bar{V}_H|_{n\textrm{ soft emissions}} &= \frac{i C_A g^2 m_H^2}{(2\pi)^4} \int \! \d^2 q_{\perp} \int \! \d \alpha \int \! \d\beta \nonumber \\
	& \qquad \qquad \qquad \times \frac{\bar{V}_H|_{n-1\textrm{ soft emissions}}}{[q^2+i\varepsilon][(q-k_1)^2+i\varepsilon][(q+k_2)^2+i\varepsilon]} \label{eq:VHsoft1}
\end{align}
where
\begin{align}
	q^\mu &= \alpha k_1^\mu + \beta k_2^\mu + q_\perp^\mu \;. \label{eq:qSudakovDecomp}
\end{align}
This diagram contains divergences both when $q$ is collinear to $k_1$ and when it is collinear to $k_2$. In order to bring the result into the form of equation~(\ref{eq:VHcollinear}) we must separate out these two regions. An effective way to accomplish this is to multiply the integrand of~(\ref{eq:VHsoft1}) by
\begin{align}
	1 &= \textrm{PV}\left( \frac{\alpha}{\alpha+\beta}  \right) +\textrm{PV}\left( \frac{\beta}{\alpha+\beta}  \right)
\end{align}
where PV denotes the Cauchy principal value. Now the term proportional to $\alpha/(\alpha+\beta)$ possesses only a collinear divergence with respect to $k_1$, whereas the other piece has only a collinear divergence when $q \propto k_2$. We may then perform the $\beta(\alpha)$ integrals in the first(second) term by contour integration. Only the piece coming from the $[q^2+i\varepsilon]$ pole is relevant to the $(1-z)^{-1}$ divergence. Keeping just this piece, we obtain
\begin{align}
	\bar{V}_H|_{n\textrm{ soft emissions} }&= -\frac{\alpha_s}{2\pi} C_A \int \! \frac{\d\bs{q}_\perp^2}{\bs{q}_\perp^2} \left(  \int_0^1 \! \d\alpha \frac{1}{\alpha +\frac{\bs{q}_\perp^2}{\alpha m_H^2}}  +  \int_0^1 \! \d|\beta| \frac{1}{|\beta| +\frac{\bs{q}_\perp^2}{|\beta| m_H^2}}  \right) \nonumber \\
	& \qquad \qquad \qquad \times \bar{V}_H|_{n-1\textrm{ soft emissions}}~.
\end{align}
Changing variables as $\alpha = 1-z$, $|\beta| = 1-z$ and noting that the momentum fraction integrals are effectively cutoff at $|\bs{q}_\perp|/m_H$, this becomes
\begin{align}
	\bar{V}_H|_{n\textrm{ soft emissions}} &= - 2 \frac{\alpha_s}{2\pi} C_A \int \! \frac{\d\bs{q}_\perp^2}{\bs{q}_\perp^2} \int_0^{1-|\bs{q}_{\perp}|/m_H} \frac{\d z}{1-z}\bar{V}_H|_{n-1\textrm{ soft emissions}}
\end{align}
and so equation~(\ref{eq:divergentSudakov}) becomes
\begin{equation}
	\bar{V}_H = T(\bs{Q}_\perp,m_H) \overline{\mathcal{M}}(gg \to H) ~,
\end{equation}
with
\begin{equation}
T(\bs{Q}_\perp,m_H) =
\exp \left(  -\int_{\bs{Q}_\perp^2}^{m_H^2} \frac{\d\bs{q}_\perp^2}{\bs{q}_\perp^2} \frac{\alpha_s(\bs{q}_\perp^2)}{2\pi} \int_0^{1-|\bs{q}_{\perp}|/m_H} \d z \left[  z P_{gg}(z) +n_f P_{qg}(z)  \right] \right)~. \label{eq:MyAllOrdersSudakov}
\end{equation}
Note that this result differs from equations (\ref{eq:DurhamSudakov})--(\ref{eq:DurhamMU}). In sections~\ref{sec:FullNLO} and \ref{sec:NLO2} we shall provide further evidence that equation (\ref{eq:MyAllOrdersSudakov}) is indeed the correct form of the Sudakov factor.

\subsubsection{The Sudakov derivative}
%.....................................................................

%%%%%%%%%%%%%%%%%%%%%%%%%%%%%%%%%%%%
\begin{figure}[b]
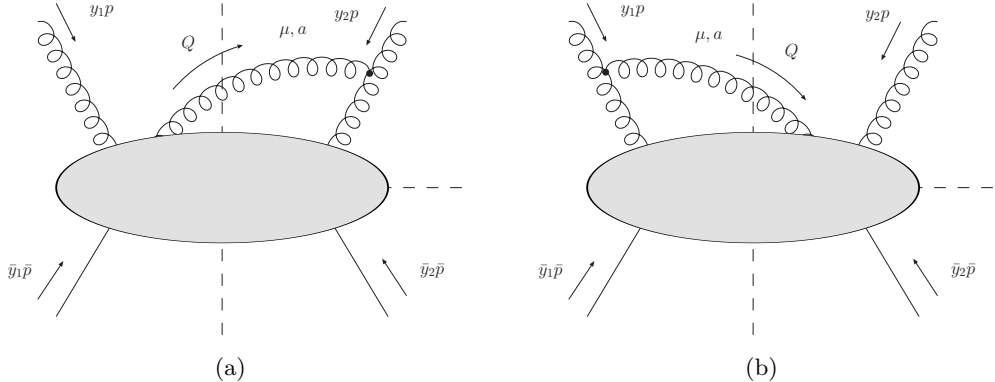

	\centering
	\subfigure[]{\label{fig:SudDeriv1}\includegraphics[width=0.4\textwidth]{FeynmanDiagrams/SoftEffects/2/2.epsi}} \hspace{0.05\textwidth}
	\subfigure[]{\label{fig:SudDeriv2}\includegraphics[width=0.4\textwidth]{FeynmanDiagrams/SoftEffects/3/3.epsi}}
	\caption{Diagrams for the last emission collinear to the upper hadron which generate the soft part of the splitting kernel.}\label{fig:SudDeriv}
\end{figure}
%%%%%%%%%%%%%%%%%%%%%%%%%%%%%%%%%%%%

We now turn our attention to a proper treatment of the (unregularised) splitting kernels entering due to the final two $s$-channel emissions. The only issue is with the $K^{gg}$ kernel, which diverges in the soft limit. In order to understand how to treat this region correctly we focus on the final emission collinear to the upper hadron, assuming all previous emissions have been collected into the pdf. 

As discussed in section~\ref{sec:CEPLO}, we may compute the amplitude by taking the cuts shown in figures~\ref{fig:CEPgeneralcuts1} and \ref{fig:CEPgeneralcuts5} only. Furthermore, if we wish to study the soft limit of the final emission, only the diagrams shown in figure~\ref{fig:SudDeriv} and the analogous diagrams with the Higgs emitted on the other side of the cut will contribute. In addition, we need only consider the sub-set of these diagrams in which $Q$ attaches to on-shell particles. The result is then given by the soft insertion rules (see for example~\cite{Bassetto:1984ik}). After summing the two diagrams in figure~\ref{fig:SudDeriv}, for example, the amplitude to the left of the cut, $\ket{L_1}^{\mu,a}$, may be written
\begin{align}
	\ket{L_1}^{\mu,a} &= \sum_{i=0}^{n+1} \left(  \frac{-g l_i^{\mu}}{l_i\cdot Q} \right) \bs{T}^a_{l_i} \ket{L_0}~,
\end{align}
where $l_0 = y_1 p$, $l_{n+1}=\bar{y}_2 \bar{p}$ and the other $l_i$ are the momenta of the particles crossing the cut. The amplitude without the soft gluon is represented by $\ket{L_0}$. Since $Q$ is the last emission collinear to the upper proton, we may take $l_i \propto \bar{p}$ for $i \neq 0$. Including the amplitude to the right of the cut, $\ket{R_1}$, using the same soft insertion formula and including an integration over the intermediate phase-space,  we obtain
\begin{align}
	A|_{\textrm{\ref{fig:SudDeriv}}} &= -\int \! \d(PS^n) \int \! \frac{\d^4 Q}{(2\pi)^3} \delta_{(+)}(Q^2) \frac{g^2 \; p\cdot \bar{p}}{p\cdot Q \; \bar{p} \cdot Q} \nonumber \\
	& \qquad \times \sum_{i=1}^{n+1} \left(   \bra{R_0}(\bs{T}_{l_i}^a)^\dagger \bs{T}_{l_0}^a \ket{L_0} +\bra{R_0}(\bs{T}_{l_0}^a)^\dagger \bs{T}_{l_i}^a \ket{L_0} \right)~,
\end{align}
where $\d(PS^n)$ is the phase-space of the cut diagram without the soft gluon. Then, using colour conservation:
\begin{align}
	\sum_{i=1}^{n+1} \bs{T}_{l_i}^a &= - \bs{T}^a_{l_0} \;, \\
	(\bs{T}_{l_0}^a)^\dagger \bs{T}_{l_0}^a &= C_A \;,
\end{align}
we find
\begin{align}
	A|_{\textrm{\ref{fig:SudDeriv}}} &= \int \! \d(PS^n) \int \! \frac{\d^4 Q}{(2\pi)^3} \delta_{(+)}(Q^2) \frac{C_A g^2 \; p\cdot \bar{p}}{p \cdot Q \; \bar{p} \cdot Q} 2 \braket{R_0|L_0} \;. \label{eq:SudDerivSoft1}
\end{align}
Now note that (\ref{eq:SudDerivSoft1}) possesses divergences when $Q$ becomes collinear to either $p$ or $\bar{p}$. As in the case of the Sudakov factor, we may separate out these regions by multiplying the integrand by $1=(\alpha+\beta)/(\alpha+\beta)$, where this time
\begin{align}
	Q^\mu &= \alpha p^\mu + \beta \bar{p}^\mu + Q_\perp^\mu \;.
\end{align}
Since we are interested in the divergences with respect to the upper hadron, we keep only the $\alpha/(\alpha+\beta)$ piece. The only effect of the phase-space on $Q$ is to introduce the constraint $\Theta(y_1 p^+ -Q^+)$. Thus, (\ref{eq:SudDerivSoft1}) becomes
\begin{align}
	A|_{\textrm{\ref{fig:SudDeriv}}} &= \frac{2 g^2 C_A}{(2\pi)^3} \int \! \frac{\d^2 \bs{Q}_\perp}{\bs{Q}_\perp^2} \int_0^{y_1} \! \d\alpha \frac{1}{\alpha + \frac{\bs{Q}_\perp^2}{4\alpha s}} \int \! \d(PS^n) \braket{R_0|L_0} \;.
\end{align}
Comparing this to equation~(\ref{eq:splittingkernels}), by taking the limit $x_i \to y_i$, we see that the correct form of the unregularised splitting kernel is
\begin{align}
	\tilde{K}^{gg}_{(0)}(x_1,x_2|y_1,y_2) &= C_A  \left[  \frac{x_1}{y_1}  \frac{x_1 \; \vartheta_{11}^0(x_1,x_1-y_1) }{\left(x_1-y_1+\frac{\bs{Q}_\perp^2}{(x_1-y_1)4s}\right)} \right. \nonumber \\
	& \quad \quad \qquad+ \frac{x_2}{y_2}  \frac{x_2 \; \vartheta_{11}^0(x_2,x_2-y_2)}{\left( x_2-y_2+\frac{\bs{Q}_\perp^2}{(x_2-y_2)4s}\right)}    \nonumber \\
	& \quad \quad \qquad+2\frac{x_1x_2+y_1y_2}{y_1y_2}\vartheta_{111}^0(x_1,-x_2,x_1-y_1) \nonumber \\
	& \quad \quad \qquad +2\frac{x_1x_2}{y_1y_2}\frac{x_1y_1+x_2y_2}{(x_1+x_2)^2}\vartheta_{11}^0(x_1,-x_2) \Bigg] \;.\label{eq:trueUnregKgg}
\end{align}
To the soft divergent pieces we may now add and subtract a term proportional to a delta-function:
\begin{align}
	\frac{x}{y} \frac{x\; \vartheta_{11}^0(x,x-y)}{\left( x-y+\frac{\bs{Q}_\perp^2}{(x-y)4s}\right)} &= \frac{x}{y}\left[ \frac{x \; \vartheta_{11}^0(x,x-y)}{\left( x-y+\frac{\bs{Q}_\perp^2}{(x-y)4s}\right)} \right. \nonumber \\
	& \qquad \quad \left. - \delta(x-y) \int \! \d x' \frac{x'\vartheta_{11}^0(x',x'-y)}{\left( x'-y+\frac{\bs{Q}_\perp^2}{(x'-y)4s}\right)} \right] \nonumber \\
	& \qquad \quad + \delta(x-y) \int \! \d x' \frac{x'\vartheta_{11}^0(x',x'-y)}{\left( x'-y+\frac{\bs{Q}_\perp^2}{(x'-y)4s}\right)} \;.
\end{align}
In the first term, contained in square brackets, we may take the $\bs{Q}_\perp \to 0$ limit since the soft region cancels, leaving us with something regularised by the plus-prescription. The integral in the second term may be done explicitly, giving a logarithm plus non-logarithmic terms which we neglect. Substituting this identity into equation~(\ref{eq:trueUnregKgg}) it becomes
\begin{align}
	\tilde{K}^{gg}(x_1,x_2|y_1,y_2) &= K^{gg}(x_1,x_2|y_1,y_2) - \frac{\alpha_s}{4\pi} C_A \delta(x_1-y_1) \ln \left(  \frac{4y_1^2s+\bs{Q}_\perp^2}{\bs{Q}_\perp^2}  \right) \nonumber \\
	& \quad - \frac{\alpha_s}{4\pi} C_A \delta(x_2-y_2) \ln \left(  \frac{4y_2^2s+\bs{Q}_\perp^2}{\bs{Q}_\perp^2}  \right) \;.
\end{align}
Applying this to the splitting kernels appearing in equation~(\ref{eq:CEPwithKtilde}) gives, e.g.
\begin{align}
	\tilde{K}^{ga}\left( \frac{x_1}{2},0 \Big| y_1,y_2 \right) &= K^{ga}\left( \frac{x_1}{2},0 \Big| y_1,y_2 \right) - \delta^{ga} \frac{\alpha_s}{2\pi} C_A \delta \left(  \xi - \frac{x_1}{2} \right) \ln\left( \frac{m_H^2}{\bs{Q}_\perp^2}  \right)~,
\end{align}
where we replaced $x_1^2 s \to m_H^2$, which is correct to logarithmic accuracy. This logarithm in the Higgs mass may then be written in terms of the derivative of the Sudakov factor:
\begin{align}
	\tilde{K}^{ga}\left( \frac{x_1}{2},0 \Big| y_1,y_2 \right) &= K^{ga}\left( \frac{x_1}{2},0 \Big| y_1,y_2 \right) \nonumber \\
	& \qquad  - \delta^{ga} \delta \left(  \xi - \frac{x_1}{2} \right) \frac{2}{\sqrt{T(\bs{Q}_\perp,m_H)}} \frac{ \partial \sqrt{T(\bs{Q}_\perp,m_H)}}{\partial \ln(\bs{Q}_\perp^2)} \;.
\end{align}
All together then, including the Sudakov factor, equation~(\ref{eq:CEPpdfderiv}) becomes
\begin{align}
	\mathcal{A}_{\textrm{CEP}} &\approx  \int \! \frac{\d\bs{Q}_\perp^2}{\bs{Q}_\perp^4}  \frac{\partial}{\partial \ln(\bs{Q}_\perp^2)} \left[ H_g\left(\frac{x_1}{2},\frac{x_1}{2};\bs{Q}_\perp^2\right) \sqrt{T(\bs{Q}_\perp,m_H)} \right] \nonumber \\
	& \quad \times \frac{\partial}{\partial \ln(\bs{Q}_\perp^2)}  \left[ H_g\left(\frac{x_2}{2},\frac{x_2}{2};\bs{Q}_\perp^2\right) \sqrt{T(\bs{Q}_\perp,m_H)} \right]  \nonumber \\
	& \quad \times \frac{\pi^3 2^4 (-i)}{x_1x_2(N^2-1)} \overline{\mathcal{M}}(gg \to H)  \;.
\end{align}
Finally, using equation~(\ref{eq:RgDeff}) and assuming that $R_g$ depends only weakly on $\bs{Q}_\perp^2$, we find
\begin{align}
	\mathcal{A}_{\textrm{CEP}} &\approx  \int \! \frac{\d\bs{Q}_\perp^2}{\bs{Q}_\perp^4}  f_g(x_1,0,\bs{Q}_\perp^2,m_H^2) f_g(x_2,0,\bs{Q}_\perp^2,m_H^2)  \nonumber \\
	& \qquad \qquad \times \frac{\pi^3 2^4 (-i)}{x_1x_2(N^2-1)} \overline{\mathcal{M}}(gg \to H)  \;.
\end{align}

%% file: Sections/ExplicitNLO.tex
\section{An explicit next-to-leading order calculation}\label{sec:FullNLO}
%%%%%%%%%%%%%%%%%%%%%%%%%%%%%%%%%%%%%%%%%%%%%%%%%%%%%%%%%%%%%%%%

Having discussed the result at all orders, we now turn to a description of our next-to-leading order calculation of the amplitude for two quarks of different flavour to scatter into two quarks and a Higgs. This calculation will serve as an explicit check of the all orders result presented in the previous section and also offers the possibility to extend the result to next-to-leading order accuracy\footnote{The Durham group do include a K-factor in their calculation of central exclusive Higgs production~\cite{Khoze:2001xm}, taken from the calculation of inclusive Higgs production at next-to-leading order~\cite{Spira:1995rr,Kunszt:1996yp}. They do not however explicitly evaluate the next-to-leading order contribution.}.  

As stated in section~\ref{sec:CEPLO}, we may calculate the amplitude from the cuts~\ref{fig:CEPgeneralcuts1}, \ref{fig:CEPgeneralcuts5}. We limit ourselves to a calculation of the virtual corrections to one side of a cut; it is the Sudakov factor we are interested in probing here and we expect only this set of diagrams to contribute to it. After presenting our results we shall comment on why the diagrams with an additional gluon crossing the cut cannot contribute to the Sudakov factor. 

\begin{figure}[h]
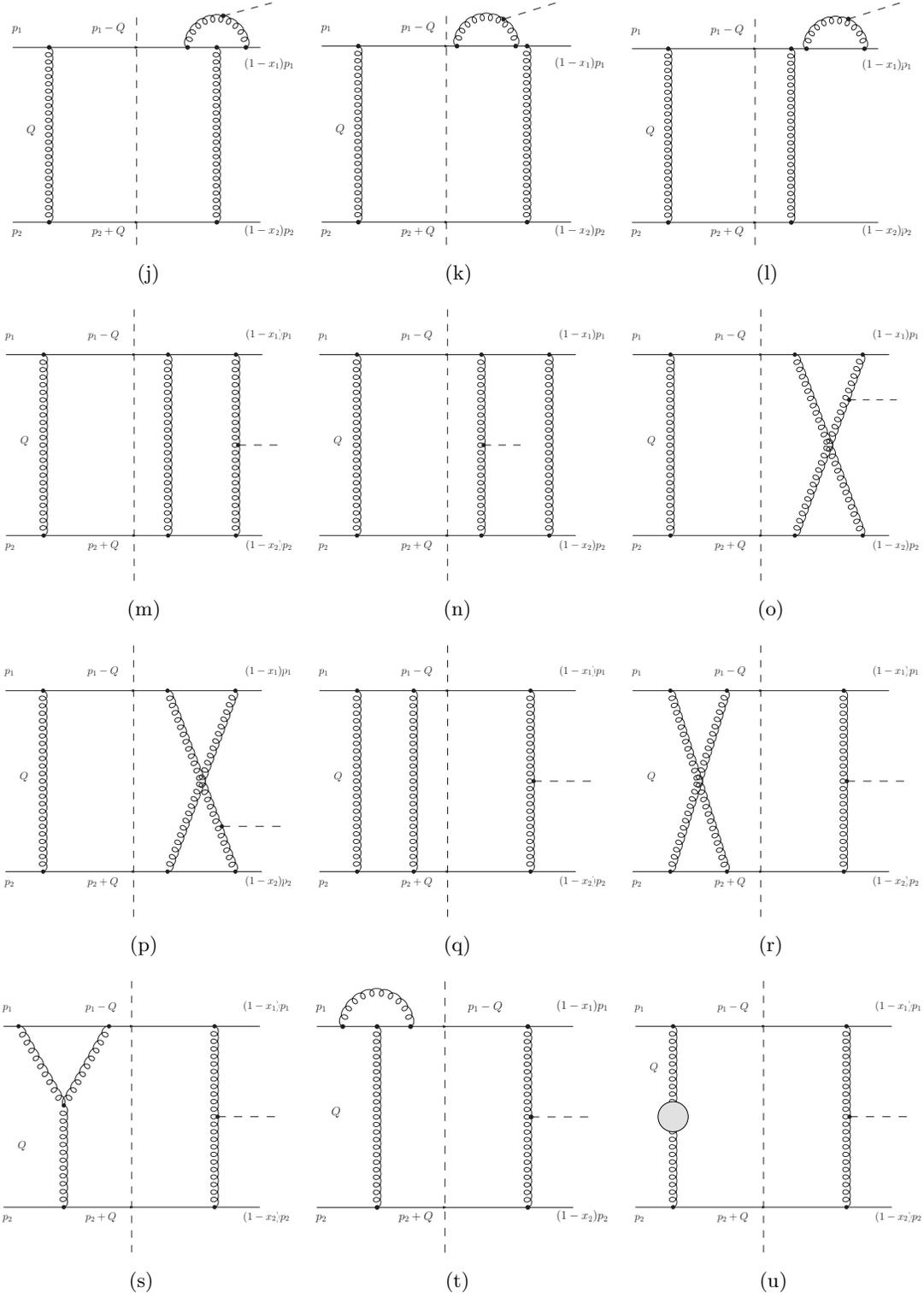

	\center
%	\subfigure[]{\label{fig:NLOa}\includegraphics[height=0.25\textwidth]{FeynmanDiagrams/CEPNLO/a/a.epsi}} \hspace{0.01\textwidth}
%	\subfigure[]{\label{fig:NLOb}\includegraphics[height=0.25\textwidth]{FeynmanDiagrams/CEPNLO/b/b.epsi}} \hspace{0.01\textwidth}
%	\subfigure[]{\label{fig:NLOc}\includegraphics[height=0.25\textwidth]{FeynmanDiagrams/CEPNLO/c/c.epsi}} \\
%	\subfigure[]{\label{fig:NLOd}\includegraphics[height=0.25\textwidth]{FeynmanDiagrams/CEPNLO/d/d.epsi}} \hspace{0.01\textwidth}
%	\subfigure[]{\label{fig:NLOe}\includegraphics[height=0.25\textwidth]{FeynmanDiagrams/CEPNLO/e/e.epsi}} \hspace{0.01\textwidth}
%	\subfigure[]{\label{fig:NLOf}\includegraphics[height=0.25\textwidth]{FeynmanDiagrams/CEPNLO/f/f.epsi}} \\
%	\subfigure[]{\label{fig:NLOg}\includegraphics[height=0.28\textwidth]{FeynmanDiagrams/CEPNLO/g/g.epsi}} %\hspace{0.01\textwidth}
%	\subfigure[]{\label{fig:NLOh}\includegraphics[height=0.28\textwidth]{FeynmanDiagrams/CEPNLO/h/h.epsi}} %\hspace{0.01\textwidth}
%	\subfigure[]{\label{fig:NLOi}\includegraphics[height=0.28\textwidth]{FeynmanDiagrams/CEPNLO/i/i.epsi}} \\
	\subfigure[]{\label{fig:NLOj}\includegraphics[height=0.25\textwidth]{FeynmanDiagrams/CEPNLO/j/j.epsi}} \hspace{0.01\textwidth}
	\subfigure[]{\label{fig:NLOk}\includegraphics[height=0.25\textwidth]{FeynmanDiagrams/CEPNLO/k/k.epsi}} \hspace{0.01\textwidth}
	\subfigure[]{\label{fig:NLOl}\includegraphics[height=0.25\textwidth]{FeynmanDiagrams/CEPNLO/l/l.epsi}} \\
	\subfigure[]{\label{fig:NLOm}\includegraphics[height=0.28\textwidth]{FeynmanDiagrams/CEPNLO/m/m.epsi}} \hspace{0.01\textwidth}
	\subfigure[]{\label{fig:NLOn}\includegraphics[height=0.28\textwidth]{FeynmanDiagrams/CEPNLO/n/n.epsi}} \hspace{0.01\textwidth}
	\subfigure[]{\label{fig:NLOo}\includegraphics[height=0.28\textwidth]{FeynmanDiagrams/CEPNLO/o/o.epsi}} \\
	\subfigure[]{\label{fig:NLOp}\includegraphics[height=0.28\textwidth]{FeynmanDiagrams/CEPNLO/p/p.epsi}} \hspace{0.01\textwidth}
	\subfigure[]{\label{fig:NLOq}\includegraphics[height=0.28\textwidth]{FeynmanDiagrams/CEPNLO/q/q.epsi}} \hspace{0.01\textwidth}
	\subfigure[]{\label{fig:NLOr}\includegraphics[height=0.28\textwidth]{FeynmanDiagrams/CEPNLO/r/r.epsi}} \\
	\subfigure[]{\label{fig:NLOs}\includegraphics[height=0.28\textwidth]{FeynmanDiagrams/CEPNLO/s/s.epsi}} \hspace{0.01\textwidth}
	\subfigure[]{\label{fig:NLOt}\includegraphics[height=0.28\textwidth]{FeynmanDiagrams/CEPNLO/t/t.epsi}} \hspace{0.01\textwidth}
	\subfigure[]{\label{fig:NLOu}\includegraphics[height=0.28\textwidth]{FeynmanDiagrams/CEPNLO/u/u.epsi}} 
	\caption{Virtual corrections contributing to the cut $qq' \to q \oplus H \oplus q'$ amplitude at next-to-leading order, in the high energy limit. Not shown are those diagrams obtained by exchanging $x_1$ and $x_2$ and those in which the Higgs is radiated on the left of the cut.}\label{fig:CEPNLO}
\end{figure}

The set of diagrams we must calculate when the Higgs is to the right of the cut are shown in figure~\ref{fig:CEPNLO}, with a similar set for the Higgs to the left of the cut not shown. All other diagrams may be obtained by exchanging $x_1$ and $x_2$. We perform the loop integrals using the techniques described in \cite{Ellis:2005zh}, which we have implemented using Mathematica~\cite{Mathematica} and FORM~\cite{FORM}. We also use the Mathematica package~FeynCalc~\cite{Mertig:1990an} to simplify the numerator algebra. All of our calculations are performed in Feynman gauge and using the large top mass effective theory described in appendix~\ref{app:largetop}.

In addition, we keep only terms not suppressed by additional powers of $x_i$ or $\bs{Q}_\perp^2$, relative to the lowest order case. The cut sets
\begin{align}
	Q^\pm &\approx \frac{\pm \bs{Q}_\perp^2}{\sqrt{2s}}
\end{align}
and we may also make the approximation
\begin{align}
	\bar{u}_{\sigma_i}(p_i') &\approx \bar{u}_{\sigma_i}(p_i) \;,
\end{align}
for the final-state quark spinor wavefunctions. This allows us to make the replacement
\begin{align}
	u_{\sigma_i}(p_i)\bar{u}_{\sigma_i}(p_i')  \to \frac{\slashed{p}_i}{2}~,
\end{align}
where we have dropped a term proportional to $\gamma^5$, which is not relevant since the amplitude is $CP$-invariant. We shall also present here only those terms either enhanced by a logarithm or divergent as $\epsilon \to 0$.

Working in dimensional regularisation, with $d=4-2\epsilon$ space-time dimensions, the results for each cut diagram are
\begin{flushleft}

\begin{align}
	C_{\textrm{NLO}}|_{\textrm{\ref{fig:NLOa}}} &= A_0(\mu) \int \! \frac{\d \bs{Q}_\perp^2}{\bs{Q}_\perp^4} \frac{C_A \alpha_s}{\pi}\left(   \frac{11 \mathcal{N}}{12 \epsilon} +\frac{11}{12} \ln \left(  \frac{\mu^2}{\bs{Q}^2_\perp} \right)  -\frac{1}{2}  \ln^2\left(  \frac{m_H^2}{\bs{Q}^2_\perp}  \right)     \right. \nonumber \\
	& \qquad \qquad  \qquad \qquad \quad+ \left(  \frac{3}{8} -i \pi \right) \ln\left(  \frac{m_H^2}{\bs{Q}^2_\perp}  \right) + \frac{m_H^2}{\bs{Q}^2_\perp} \left[  \frac{7 \mathcal{N}}{24 \epsilon}   \right. \nonumber \\
	& \qquad \qquad \qquad \qquad \quad  \left. \left. +\frac{7}{24}\ln\left(  \frac{\mu^2}{m_H^2}  \right)  -\frac{7 i \pi}{24} + \frac{49}{144} \right]  \right)\\ \nonumber \\
	C_{\textrm{NLO}}|_{\textrm{\ref{fig:NLOb}}} &= A_0(\mu) \int \! \frac{\d \bs{Q}_\perp^2}{\bs{Q}_\perp^4} \frac{C_A \alpha_s}{\pi}\left(   -\frac{19 \mathcal{N}}{48 \epsilon} - \frac{19}{48}\ln\left(  \frac{\mu^2}{\bs{Q}_\perp^2}  \right)  \right) \\ \nonumber \\
	C_{\textrm{NLO}}|_{\textrm{\ref{fig:NLOc}}} &= A_0(\mu) \int \! \frac{\d \bs{Q}_\perp^2}{\bs{Q}_\perp^4} \frac{C_A \alpha_s}{\pi} \frac{m_H^2}{\bs{Q}_\perp^2}  \left(   -\frac{7\mathcal{N}}{24\epsilon} - \frac{7}{24}\ln\left(  \frac{\mu^2}{m_H^2}  \right) +\frac{7 i\pi}{24} - \frac{49}{144} \right)  %\\ \nonumber \\
\end{align}
%%%%%%%%%%%%%%%%%%
\begin{align}
	C_{\textrm{NLO}}|_{\textrm{\ref{fig:NLOd}}} &= A_0(\mu) \int \! \frac{\d \bs{Q}_\perp^2}{\bs{Q}_\perp^4} \frac{C_A \alpha_s}{\pi}\left(   -\frac{\mathcal{N}}{8\epsilon} -\frac{1}{8} \ln\left( \frac{\mu^2}{\bs{Q}_\perp^2}\right) \right) \\ \nonumber \\
	C_{\textrm{NLO}}|_{\textrm{\ref{fig:NLOe}}} &= A_0(\mu) \int \! \frac{\d \bs{Q}_\perp^2}{\bs{Q}_\perp^4} (C_A-2C_F)\frac{\alpha_s}{\pi}\left(   \frac{\mathcal{N}}{4\epsilon^2}\left(  \frac{\mu^2}{\bs{Q}^2_\perp} \right)^\epsilon +\frac{3\mathcal{N}}{8\epsilon} + \frac{3}{8}\ln\left( \frac{\mu^2}{\bs{Q}^2_\perp} \right) \right) \\ \nonumber \\
	C_{\textrm{NLO}}|_{\textrm{\ref{fig:NLOf}}} &= A_0(\mu) \int \! \frac{\d \bs{Q}_\perp^2}{\bs{Q}_\perp^4} (5 C_A -2n_f)\frac{\alpha_s}{\pi}\left(   \frac{\mathcal{N}}{12\epsilon} + \frac{1}{12}\ln\left( \frac{\mu^2}{\bs{Q}^2_\perp}\right) \right) 
%  \\ \nonumber \\
%
\end{align}
%%%%%%%%%%%%%%%%%%
\begin{align}
	C_{\textrm{NLO}}|_{\textrm{\ref{fig:NLOg}}} &= A_0(\mu) \int \! \frac{\d \bs{Q}_\perp^2}{\bs{Q}_\perp^4} \left(   \frac{C_A \alpha_s}{32 \pi}\frac{\mathcal{N}}{\epsilon} + \frac{C_A \alpha_s}{32 \pi}\ln\left( \frac{\mu^2}{m_H^2} \right) +\frac{C_A \alpha_s}{8\pi} \ln^2\left(  \frac{m_H^2}{\bs{Q}_\perp^2}\right) \right. \nonumber \\
	&\qquad \qquad \qquad \qquad \quad +\frac{C_A \alpha_s}{16 \pi}(4 i \pi -1)\ln \left( \frac{m_H^2}{\bs{Q}_\perp^2}\right) + \frac{C_A \alpha_s}{16 \pi}\ln(x_1)  \nonumber \\
	&\qquad \qquad \qquad \qquad \quad \left. + \frac{\epsilon_G(Q^2)}{2}\left(  \ln\left( \frac{x_2 s}{\bs{Q}^2_\perp} \right) +i\pi \right) \right) \\ \nonumber \\
	C_{\textrm{NLO}}|_{\textrm{\ref{fig:NLOh}}} &= A_0(\mu) \int \! \frac{\d \bs{Q}_\perp^2}{\bs{Q}_\perp^4} \left(   \frac{C_A \alpha_s}{32 \pi}\frac{\mathcal{N}}{\epsilon} + \frac{C_A \alpha_s}{32 \pi}\ln\left( \frac{\mu^2}{m_H^2} \right) +\frac{C_A \alpha_s}{8\pi} \ln^2\left(  \frac{m_H^2}{\bs{Q}_\perp^2}\right) \right. \nonumber \\
	&\qquad \qquad \qquad \qquad \quad +\frac{C_A \alpha_s}{16 \pi}(4 i \pi -1)\ln \left( \frac{m_H^2}{\bs{Q}_\perp^2}\right) + \frac{C_A \alpha_s}{16 \pi}\ln(x_1)  \nonumber \\
	&\qquad \qquad \qquad \qquad \quad \left. + \frac{\epsilon_G(Q^2)}{2}\ln\left( \frac{x_2 s}{\bs{Q}^2_\perp} \right)  \right) \\ \nonumber \\
	C_{\textrm{NLO}}|_{\textrm{\ref{fig:NLOi}}} &= A_0(\mu) \int \! \frac{\d \bs{Q}_\perp^2}{\bs{Q}_\perp^4}   \frac{C_A \alpha_s}{\pi} \left(	\frac{\mathcal{N}}{8 \epsilon} +\frac{1}{8} \ln \left(  \frac{\mu^2}{\bs{Q}^2_\perp} \right) \right) \\ \nonumber \\
	C_{\textrm{NLO}}|_{\textrm{\ref{fig:NLOj}}} &= A_0(\mu) \int \! \frac{\d \bs{Q}_\perp^2}{\bs{Q}_\perp^4} (C_A-2C_F) \frac{\alpha_s}{8 \pi} \ln \left( \frac{1}{x_1}  \right) \\ \nonumber \\
	C_{\textrm{NLO}}|_{\textrm{\ref{fig:NLOk}}} &= C_{\textrm{NLO}}|_{\textrm{\ref{fig:NLOl}}} =A_0(\mu) \int \! \frac{\d \bs{Q}_\perp^2}{\bs{Q}_\perp^4}  \frac{C_F\alpha_s}{8 \pi} \ln \left( \frac{1}{x_1}  \right) \\ \nonumber \\
	C_{\textrm{NLO}}|_{\textrm{\ref{fig:NLOm}}} &= A_0(\mu) \int \! \frac{\d \bs{Q}_\perp^2}{\bs{Q}_\perp^4}  \left(-\frac{1}{N}\right) \left(  \frac{\alpha_s}{2 \pi}\frac{\mathcal{N}}{\epsilon^2} \left(  \frac{\mu^2}{\bs{Q}_\perp^2}  \right)^\epsilon  +\frac{\alpha_s}{4\pi}\ln^2\left( \frac{m_H^2}{\bs{Q}_\perp^2} \right)  \right. \nonumber \\
	& \qquad \qquad \qquad \qquad \quad \left. +  \frac{\alpha_s}{2 \pi} i\pi \ln \left(  \frac{m_H^2}{\bs{Q}_\perp^2}  \right)  +  \frac{\epsilon_G(Q^2)}{C_A} \left( \ln\left(\frac{m_H^2}{\bs{Q}_\perp^2} \right) + i\pi \right) \right) \\ \nonumber \\
	C_{\textrm{NLO}}|_{\textrm{\ref{fig:NLOn}}} &= C_{\textrm{NLO}}|_{\textrm{\ref{fig:NLOm}}}+ A_0(\mu) \int \! \frac{\d \bs{Q}_\perp^2}{\bs{Q}_\perp^4}\frac{i \pi}{N}\frac{\alpha_s}{\pi} \left(  \frac{\mathcal{N}}{2\epsilon} + \frac{1}{2} \ln\left( \frac{\mu^2}{\bs{Q}_\perp^2}  \right)  \right)  	\\ \nonumber \\
	C_{\textrm{NLO}}|_{\textrm{\ref{fig:NLOo}}} &= C_{\textrm{NLO}}|_{\textrm{\ref{fig:NLOp}}}= A_0(\mu) \int \! \frac{\d \bs{Q}_\perp^2}{\bs{Q}_\perp^4}  \left(\frac{1}{N}-\frac{C_A}{2} \right) \left(  \frac{\alpha_s}{2\pi}\frac{\mathcal{N}}{\epsilon^2} \left( \frac{\mu^2}{\bs{Q}_\perp^2} \right)^\epsilon  \right. \nonumber \\ 
	& \qquad \qquad \qquad \qquad \quad  + \frac{\alpha_s}{4\pi}\ln^2\left( \frac{m_H^2}{\bs{Q}_\perp^2}  \right)  + \frac{\alpha_s}{2\pi} i\pi \ln\left( \frac{m_H^2}{\bs{Q}_\perp^2} \right) \nonumber \\
	& \qquad \qquad \qquad \qquad \quad \left. +\frac{\epsilon_G(Q^2)}{C_A} \left(  \ln\left( \frac{m_H^2}{\bs{Q}_\perp^2} \right) +i \pi  \right) \right)  %\\ \nonumber \\
\end{align}
%%%%%%%%%%%%%%%%%%%%%
\begin{align}
	C_{\textrm{NLO}}|_{\textrm{\ref{fig:NLOq}}} &= A_0(\mu) \int \! \frac{\d \bs{Q}_\perp^2}{\bs{Q}_\perp^4}  \frac{1}{N} \left(   -\frac{\alpha_s}{\pi}\frac{\mathcal{N}}{\epsilon^2} \left( \frac{\mu^2}{\bs{Q}_\perp^2}   \right)^\epsilon  + \frac{2\epsilon_G(Q^2)}{C_A} \left( \ln\left( \frac{s}{\bs{Q}_\perp^2}  \right)  - i\pi \right) \right) \\ \nonumber \\
	C_{\textrm{NLO}}|_{\textrm{\ref{fig:NLOr}}} &= A_0(\mu) \int \! \frac{\d \bs{Q}_\perp^2}{\bs{Q}_\perp^4} \left( \frac{C_A}{2} - \frac{1}{N} \right) \left(   -\frac{\alpha_s}{\pi}\frac{\mathcal{N}}{\epsilon^2} \left( \frac{\mu^2}{\bs{Q}_\perp^2}   \right)^\epsilon  + \frac{2\epsilon_G(Q^2)}{C_A} \ln\left( \frac{s}{\bs{Q}_\perp^2}  \right) \right)  \\ \nonumber \\
	C_{\textrm{NLO}}|_{\textrm{\ref{fig:NLOs}}} &= C_{\textrm{NLO}}|_{\textrm{\ref{fig:NLOd}}} \\ \nonumber \\
	C_{\textrm{NLO}}|_{\textrm{\ref{fig:NLOt}}} &= C_{\textrm{NLO}}|_{\textrm{\ref{fig:NLOe}}} \\ \nonumber \\	 
	C_{\textrm{NLO}}|_{\textrm{\ref{fig:NLOu}}} &= C_{\textrm{NLO}}|_{\textrm{\ref{fig:NLOf}}}	 
\end{align}
where the usual $\overline{\textrm{MS}}$ factor is given by
\begin{align}
	\mathcal{N} &= \textrm{exp}\left[  \epsilon(-\gamma_E +\ln(4\pi))  \right]
\end{align}
and $\gamma_E$ is the Euler-Mascheroni constant. The full set of counter terms for the diagrams with the Higgs to the right of the cut give
\begin{align}
	C_{\textrm{NLO}}|_{\textrm{counter terms}} &= A_0(\mu) \int \! \frac{\d \bs{Q}_\perp^2}{\bs{Q}_\perp^4} \left( -3 \beta_0 \frac{\alpha_s}{\pi} \frac{\mathcal{N}}{\epsilon} \right) \;,
\end{align}
where $\beta_0$ is the first component of the QCD beta-function and is given by
\begin{align}
	\beta_0 &= \frac{11C_A-4T_Fn_f}{12} \;.
\end{align}

\end{flushleft}

Collecting these results together and including those related by interchanging $x_1 \leftrightarrow x_2$ and those generated by the diagrams with the Higgs to the left of the cut, we obtain
\begin{align}
	A_{\textrm{NLO}} &= A_0(\mu) \int \! \frac{\d \bs{Q}_\perp^2}{\bs{Q}_\perp^4} \left( -2 C_F \frac{\alpha_s}{\pi} \frac{\mathcal{N}}{\epsilon^2} \left( \frac{\mu^2}{\bs{Q}_\perp^2}  \right)^\epsilon - 3 C_F \frac{\alpha_s}{\pi} \frac{\mathcal{N}}{\epsilon} \left( \frac{\mu^2}{\bs{Q}_\perp^2}  \right)^\epsilon \right. \nonumber \\
	& \qquad \qquad \qquad \qquad \quad  - C_A \frac{\alpha_s}{4\pi} \ln^2\left( \frac{m_H^2}{\bs{Q}_\perp^2}  \right) + 3 \beta_0 \frac{\alpha_s}{\pi} \ln\left(  \frac{\mu^2}{\bs{Q}_\perp^2} \right)  \nonumber \\
	&\qquad \qquad \qquad \qquad \quad \left. + 2 \epsilon_G(Q^2) \ln\left( \frac{s}{\bs{Q}_\perp^2}  \right) \right) \;.
\end{align}
The $1/\epsilon$ poles here are due to collinear and soft divergences and may be written in terms of the quark-quark splitting function using the following identity
\begin{align}
	C_F\left(\frac{(\bs{Q}^2_\perp)^{-\epsilon}}{\epsilon^2} + \frac{3}{2} \frac{(\bs{Q}^2_\perp)^{-\epsilon}}{\epsilon} \right) \approx \int_0^{\bs{Q}_\perp^2} \! \frac{\d q_\perp^2}{(q_\perp^2)^{1+\epsilon}} \int_0^{1-q_\perp/|\bs{Q}_\perp|} \d z \; P_{qq}(z)~,
\end{align}
which holds up to terms not involving either $1/\epsilon$ poles or logarithms of $\bs{Q}_\perp$. Each of these factors, proportional to $P_{qq}$, is due to a parton becoming collinear with one of the on-mass-shell, external quark lines. The (final state) contributions associated with the cut quark lines will cancel with an equal and opposite term coming from final-state divergences associated with a gluon emitted across the cut. If we take $\mu^2 = \bs{Q}_\perp^2$, then the remaining factors of $P_{qq}$ are simply the virtual contribution to the expansion of the pdf, $H_q(\xi,\eta;\bs{Q}_\perp^2)$. Finally, we see the factor associated with the reggeisation of the gluon, written in terms of the gluon Regge trajectory (see equations~(\ref{eq:ReggeReplacement}) and (\ref{eq:GluonRegge})). This has precisely the form predicted in equation~(\ref{eq:ReggeReplacement}) and is also expected to factorise into the unintegrated gluon pdf.

What remains then, after accounting for all of these pieces, must be the $\mathcal{O}(\alpha_s)$ expansion of the Sudakov factor:
\begin{align}
	A_{\textrm{NLO}}|_{\textrm{Sudakov}} &=  \int \! \frac{\d \bs{Q}_\perp^2}{\bs{Q}_\perp^4} A_0(\bs{Q}_\perp)  \left( -C_A \frac{\alpha_s(\bs{Q}_\perp^2)}{4\pi} \ln^2\left(  \frac{m_H^2}{\bs{Q}_\perp^2}  \right)   \right) \;. \label{eq:MySudakovUnevolved}
\end{align}
It is interesting to see which diagrams this double logarithm derives from. If we simply took the vertex correction of figure~\ref{fig:NLOa}, we would obtain twice the double logarithmic contribution of the full answer, which is consistent with our discussion of the corrections to $\bar{V}_H$ in section~\ref{sec:HardCollinear}. This additional double logarithm is cancelled when we include the diagrams \ref{fig:NLOg}, \ref{fig:NLOh} and \ref{fig:NLOm}--\ref{fig:NLOp} (and those related by $x_1 \leftrightarrow x_2$). It is also the sum of these diagrams which generates the reggeisation of the gluons to the right of the cut. 

In order to compare equation~(\ref{eq:MySudakovUnevolved}) to the Durham result, we must first exchange $A_0(\bs{Q}_\perp)$ for $A_0(m_H)$. This may be accomplished using the leading order coupling constant evolution
\begin{align}
	\alpha_s(\mu_1^2) &= \alpha_s(\mu_2^2) \; \textrm{exp} \left[  - \int_{\mu_2^2}^{\mu_1^2} \frac{\d k_\perp^2}{k_\perp^2} \frac{\alpha_s(k_\perp^2)}{\pi} \beta_0  \right]
\end{align}
which implies at next-to-leading order
\begin{align}
	A_0(\bs{Q}_\perp) &= A_0(m_H) \left(  1 + \int_{\bs{Q}_\perp^2}^{m_H^2} \frac{\d k_\perp^2}{k_\perp^2} \frac{\alpha_s(k_\perp^2)}{\pi} \beta_0    \right)
\end{align}
and thus
\begin{align}
	A_{\textrm{NLO}}|_{\textrm{Sudakov}} &= \int \! \frac{\d \bs{Q}_\perp^2}{\bs{Q}_\perp^4} A_0(m_H)  \left( -C_A \frac{\alpha_s(\bs{Q}_\perp^2)}{4\pi} \ln^2\left(  \frac{m_H^2}{\bs{Q}_\perp^2}  \right)   + \int_{\bs{Q}_\perp^2}^{m_H^2} \frac{\d k_\perp^2}{k_\perp^2} \frac{\alpha_s(k_\perp^2)}{\pi} \beta_0  \right) \nonumber \\
	& \approx \int \! \frac{\d \bs{Q}_\perp^2}{\bs{Q}_\perp^4} A_0(m_H)  \ln \left(  T(\bs{Q}_\perp, m_H)  \right) \;.\label{eq:MyFixedOrderSudakov}
\end{align}
This is exactly the next-to-leading order term one would obtain by expanding out the Sudakov factor (with $\Delta=k_\perp/m_H$) to this order in perturbation theory.

In obtaining this result we have neglected the diagrams in which a gluon is emitted across the cut. However, we now argue that such terms cannot possibly contribute to the Sudakov factor. 

There are three regions for these diagrams which could potentially produce a logarithm in $m_H^2/\bs{Q}_\perp^2$. First are the BFKL corrections. However, these are summed into the unintegrated pdfs and so cannot contribute to the Sudakov factor. Secondly, we have hard collinear emission. This was fully accounted for in section~\ref{sec:HardCollinear}, where we saw that the only large logarithms generated by $s$-channel emissions in this region had transverse momentum less than $\bs{Q}_\perp$ and so do not form part of the Sudakov factor. Finally then, we have soft emission. To be included in the Sudakov factor, the additional soft gluon must have a larger transverse momentum than one of the quarks. However, this means it is one of the final two $s$-channel emissions making up $C^{qq}_{\textrm{2 s-channel}}$. As shown in section~\ref{sec:softCEPeffects}, these soft emissions generate the Sudakov derivative.

With these considerations then, we conclude that equation~(\ref{eq:MyFixedOrderSudakov}) gives the full contribution to the Sudakov factor at this order in perturbation theory. Moreover, we have confirmed our previous result for the resummed Sudakov factor, equation~(\ref{eq:MyAllOrdersSudakov}). 

%%%%%%%%%%%%%%%%%%%%%%%%%%%%%%%%%%%%%%%%%%%%%%%%%%
\begin{figure}[h]
	\center
	\subfigure[]{\label{fig:NLOa}\includegraphics[height=0.28\textwidth]{FeynmanDiagrams/CEPNLO/a/a.epsi}} \hspace{0.01\textwidth}
	\subfigure[]{\label{fig:NLOb}\includegraphics[height=0.28\textwidth]{FeynmanDiagrams/CEPNLO/b/b.epsi}} \hspace{0.01\textwidth}
	\subfigure[]{\label{fig:NLOc}\includegraphics[height=0.28\textwidth]{FeynmanDiagrams/CEPNLO/c/c.epsi}} \\
	\subfigure[]{\label{fig:NLOd}\includegraphics[height=0.28\textwidth]{FeynmanDiagrams/CEPNLO/d/d.epsi}} \hspace{0.01\textwidth}
	\subfigure[]{\label{fig:NLOe}\includegraphics[height=0.28\textwidth]{FeynmanDiagrams/CEPNLO/e/e.epsi}} \hspace{0.01\textwidth}
	\subfigure[]{\label{fig:NLOf}\includegraphics[height=0.28\textwidth]{FeynmanDiagrams/CEPNLO/f/f.epsi}} \\
	\subfigure[]{\label{fig:NLOg}\includegraphics[height=0.28\textwidth]{FeynmanDiagrams/CEPNLO/g/g.epsi}} \hspace{0.01\textwidth}
	\subfigure[]{\label{fig:NLOh}\includegraphics[height=0.28\textwidth]{FeynmanDiagrams/CEPNLO/h/h.epsi}} \hspace{0.01\textwidth}
	\subfigure[]{\label{fig:NLOi}\includegraphics[height=0.28\textwidth]{FeynmanDiagrams/CEPNLO/i/i.epsi}} \\
\end{figure}
%%%%%%%%%%%%%%%%%%%%%%%%%%%%%%%%%%%%%%%%%%%%%%%%%%
\clearpage
%%%%%%%%%%%%%%%%%%%%%%%%%%%%%%%%%%%%%%%%%%%%%%%%%%
%%%%%%%%%%%%%%%%%%%%%%%%%%%%%%%%%%%%%%%%%%%%%%%%%%
\clearpage

%% file: Sections/Recalculation.tex
\section{A further cross-check}\label{sec:NLO2}
%%%%%%%%%%%%%%%%%%%%%%%%%%%%%%%%%%%%%%%%%%%%%%%%%%%%%%%%%%%%%%

\begin{figure}[tbp]
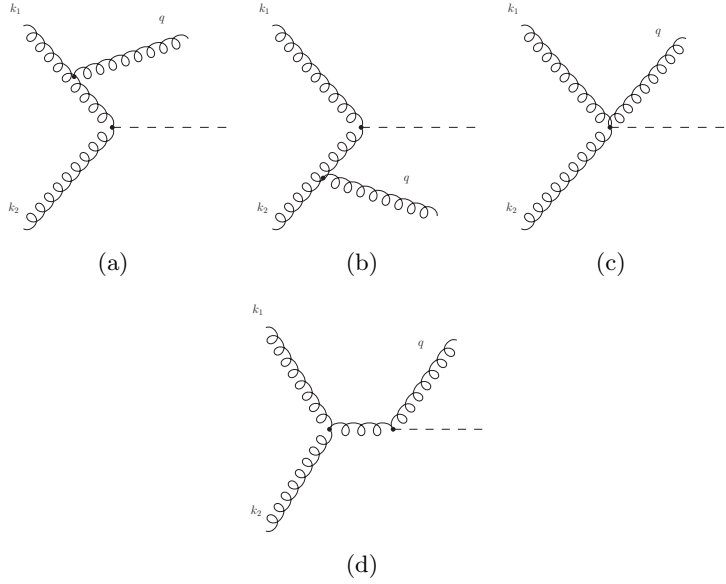

	\center
	\subfigure[]{\label{fig:Ryskin1}\includegraphics[height=0.2\textwidth]{FeynmanDiagrams/RyskinDiagrams/1/1.epsi}}  \hspace{0.01\textwidth}
	\subfigure[]{\label{fig:Ryskin2}\includegraphics[height=0.2\textwidth]{FeynmanDiagrams/RyskinDiagrams/2/2.epsi}}  \hspace{0.01\textwidth}
	\subfigure[]{\label{fig:Ryskin3}\includegraphics[height=0.2\textwidth]{FeynmanDiagrams/RyskinDiagrams/3/3.epsi}}  \\
	\subfigure[]{\label{fig:Ryskin4}\includegraphics[height=0.2\textwidth]{FeynmanDiagrams/RyskinDiagrams/4/4.epsi}}   
	\caption{Diagrams contributing to the process $gg\to Hg$.} \label{fig:RyskinDiagrams}
\end{figure}

We now discuss a re-calculation of the soft part of the Sudakov factor, which exploits the Bloch-Nordsieck theorem~\cite{Bloch:1937pw}. It serves as a further check on our earlier results. In particular, we infer the virtual corrections by considering the set of real emission diagrams making up the $gg\to gH$ amplitude, as shown in figure~\ref{fig:RyskinDiagrams}. The relevant cross-section is given by
\begin{align}
	\sigma &= \int \! \d\xi_1 \int \! \d\xi_2 \; g(\xi_1;\mu_F^2) g(\xi_2;\mu_F^2) \; \frac{1}{2\hat{s}} \frac{\hat{s}}{2} \int \! \d\alpha \int \! \d\beta \int \! \frac{\d^2q_\perp}{(2\pi)^3} \; \delta_{(+)}(q^2) \nonumber \\
	& \qquad \qquad \times (2\pi) \delta_{(+)}((k_1+k_2-q)^2 - m_H^2) |\mathcal{M}|^2
\end{align}
where the $\xi_i$ denote the momentum fractions of the incoming gluons and we parametrise the final-state gluon momentum, $q$, in terms of Sudakov variables as in equation~(\ref{eq:qSudakovDecomp}), with
\begin{align}
	\hat{s} &= 2k_1\cdot k_2  \; .
\end{align}	
The amplitude, $\mathcal{M}$, is summed over equal incoming helicities by contraction with a polarisation vector (in the $k_1$-$k_2$ centre-of-mass frame), $e^\mu = (0,\cos \phi, \sin \phi,0)$ for the incoming gluons, followed by an average over $\phi$ i.e.
\begin{align}
	\mathcal{M} &= \int_0^{2\pi}\frac{d \phi}{2\pi} e^\mu e^\nu \mathcal{M}_{\mu\nu} \;.
\end{align}

%%%%%%%%%%%%%%%%%%%%%%%%%%%%%%%%%%%%%%%%%%%%%%%%%%
%%%%%%%%%%%%%%%%%%%%%%%%%%%%%%%%%%%%%%%%%%%%%%%%%%

Next, we divide the phase-space of the emitted gluon into the regions $\alpha>\beta$ and $\alpha<\beta$, which is equivalent to dividing the phase-space about zero rapidity. We shall calculate only the contribution from the $\alpha>\beta$ region, however it will be clear that the $\alpha<\beta$ region gives an equal result. In addition, we sum over final state polarisations using\footnote{Note that, in processes involving two or more external gluons, the replacement of the gluon polarisation sum with $-g^{\mu\nu}$ is not in general valid~\cite{Cutler:1977qm}.}
\begin{align}
	\sum_\lambda \varepsilon_\lambda^\rho(k) \varepsilon^{*\rho'}_\lambda(k) &= -g^{\rho\rho'} + \frac{k_1^\rho k^{\rho'} + k_1^{\rho'}k^{\rho}}{k_1\cdot k} \;.
\end{align}
We also average over incoming gluon colours, though we stress that changing the treatment of colour in the process affects only the normalisation, since each diagram has identical colour structure. After a little algebra, which is performed using FORM, we find
\begin{align}
	\sigma|_{\alpha>\beta} &=   \int \! \d\xi_1 \int \! \d\xi_2 \; g(\xi_1;\mu_F^2) g(\xi_2;\mu_F^2) \frac{1}{2\hat{s}}   \int \! \frac{\d^2 q_\perp}{(2\pi)^2}  \int_{\frac{q_\perp}{\sqrt{\hat{s}}}}^1 \d\alpha  \; |\mathcal{M}^{\textrm{LO}}|^2 \;  C_A \; g^2 \nonumber \\ 
	& \quad \times \delta_{(+)}((k_1+k_2-q)^2-m_H^2) \left\{   \frac{1}{q_\perp^2}\left[ -4 +\frac{2}{\alpha} +4\alpha -2 \alpha^2 +\frac{\alpha^3}{2}  \right]  \right. \nonumber \\ 
	& \quad+ \frac{2}{\hat{s}}\left[  -3 +\frac{4}{\alpha} -\frac{2}{\alpha^2} +\alpha   \right]   + \frac{ q_\perp^2}{\hat{s}^2} \left[  \frac{3}{\alpha} - \frac{6}{\alpha^2} +\frac{4}{\alpha^3}  \right] \nonumber \\  
	& \left. \quad+ \frac{2q_\perp^4}{\hat{s}^3}\left[  \frac{1}{\alpha^3}-\frac{1}{\alpha^4} \right]   + \frac{ q_\perp^6}{2\hat{s}^4} \frac{1}{\alpha^5}  \right\}	  
\end{align}
where
\begin{align}
	\mathcal{M}^{\textrm{LO}} &= \int_0^{2\pi} \frac{\d\phi}{2\pi} e^\mu e^\nu \mathcal{M}^{\textrm{LO}}_{\mu\nu}
\end{align}
is the lowest order $gg\to H$ amplitude summed over equal incoming helicities. The key point to note is that only the term proportional to $1/q_\perp^2$ can generate a logarithm or a constant after the $\alpha$ integral: The remaining terms are insufficiently singular in the soft ($\alpha \to 0$) limit.  Keeping only this term then, changing variables as $\alpha =1-z$, approximating the delta-function and including the region $\alpha<\beta$, we find
\begin{align}
	\sigma &=  \int \! \d\xi_1 \int \! \d\xi_2 \; g(\xi_1;\mu_F^2) g(\xi_2;\mu_F^2) \frac{1}{2\hat{s}}   \int \! \frac{\d q_\perp^2}{q_\perp^2}  \int^{1-\frac{q_\perp}{m_H}}_0 \d z   \nonumber \\
	& \quad \times  (2\pi)\delta_{(+)}(z \hat{s}-m_H^2) |\mathcal{M}^{\textrm{LO}}|^2 \frac{\alpha_s}{\pi}\left(  z P_{gg}(z)  + \frac{C_A}{2}( 3z^3-5z^2+5z-3) )   \right) \;, \label{eq:JeffsRyskinResult}
\end{align}
where we replaced $\sqrt{\hat{s}} \to m_H$ in the upper limit of the $z$ integral, as appropriate in the soft limit. The additional $z$-dependent piece, not proportional to $P_{gg}(z)$, arises due to the restricted sum over incoming helicities and it vanishes in the soft $z \to 1$ limit.\footnote{If we had summed over all incoming helicity configurations this term would of course have been absent.}
Thus, in the soft limit, we find
\begin{align}
	\sigma|_{\textrm{soft}} \propto \int \! \frac{\d q_\perp^2}{q_\perp^2} \frac{C_A \alpha_s}{\pi} \ln \left(  \frac{m_H}{q_\perp} \right) ~,
\end{align}
which further supports the validity of equation~(\ref{eq:MyAllOrdersSudakov}).

%% file: Sections/Impact.tex
\section{Cross-section predictions}\label{sec:Pheno}
%%%%%%%%%%%%%%%%%%%%%%%%%%%%%%%%%%%%%%%%%%%%%%%%%%%%%%%%%%%%%%

\begin{figure}[tb]
	\centering
	\includegraphics[width=0.8\textwidth]{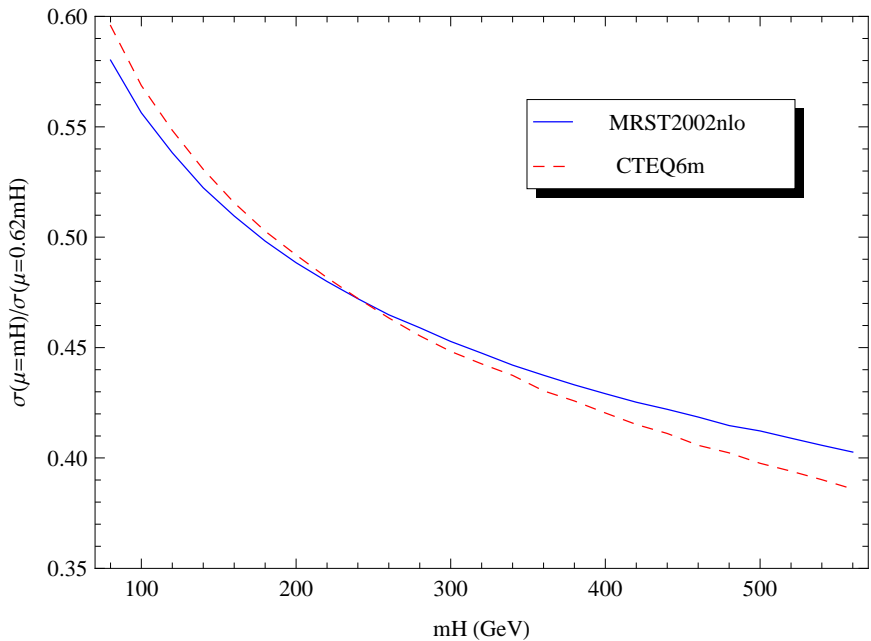}
	\caption{Ratio of the cross-section for central exclusive Higgs production at the LHC evaluated with the scale in the Sudakov factor set to $\mu=m_H$ divided by the cross-section with the scale set to $\mu=0.62 m_H$, plotted as a function of the Higgs mass. The solid blue and dashed red lines were generated using MRST2002nlo~\cite{MRST} and CTEQ6m~\cite{CTEQ} parton distributions respectively.} \label{fig:AndysNumbers}
\end{figure}

We can assess the impact that our modification of the Sudakov factor has on predictions of the central exclusive cross-section. Taking, as an example, the cross-section for central exclusive Higgs production at the LHC, with 14~TeV centre-of-mass energy we compute the cross-section, using the ExHuME Monte Carlo generator~\cite{Monk:2005ji}, placing no cuts on the final-state particles. The results are shown in figure~\ref{fig:AndysNumbers}, for two different parton distribution functions \cite{MRST,CTEQ}. We observe a suppression of the cross-section, relative to the predictions of the Durham group, by a factor $\sim 2$ which increases with increasing Higgs mass. 

%%%%%%%%%%%%%%%%%%%%%%%%%%%%%%%
%%%%%%%%%%%%%%%%%%%%%%%%%%%%%%%

%% file: Sections/Conclusions.tex
\section{Conclusions}\label{sec:conclusions}
%%%%%%%%%%%%%%%%%%%%%%%%%%%%%%%%%%%%%%%%%%%%%%

In this paper we have studied the central exclusive production process in some detail. In particular, we studied the cross-section for Higgs boson production, using QCD perturbation theory. We largely confirm the calculation previously performed by the Durham group (e.g. see \cite{Kaidalov:2003ys}), except that we disagree as to the precise form of the Sudakov factor which enters. Using the Sudakov factor that we propose leads to a suppression of the central exclusive production cross-section at the LHC by approximately a factor of two relative to the earlier predictions, for Higgs boson masses in the range 100--500~GeV.

As a point of further study, it would be interesting to assess the impact on predictions for other processes and in particular on the central exclusive production of dijets at the Tevatron, for which data exist. We do not expect to find any disagreement with the data. In fact, the reduced suppression at lower masses suggests that agreement with the data may even be slightly improved. However, one must always remember that the theoretical uncertainty on other parts of the calculation, for example the gap survival factor and unintegrated pdfs, is expected to be comparable in size to the effect induced by the change in the Sudakov factor that we have been focussing on (e.g. see ~\cite{Forshaw:2005qp}).  

We note that the fixed-order corrections we have computed form a subset of the full next-to-leading order corrections to central exclusive Higgs production, offering the possibility of extending the theoretical description of the process to this order.

\section*{Acknowledgements}
We thank Mrinal Dasgupta, Valery Khoze, Misha Ryskin and Mike Seymour for some very interesting discussions and Andy Pilkington for his assistance in producing the plot shown in section~\ref{sec:Pheno}. Thanks also to the UK Science and Technology Facilities Council (STFC) for financial support.

%% file: Sections/Appendix.tex
\section{Large top mass effective theory}\label{app:largetop}
%%%%%%%%%%%%%%%%%%%%%%%%%%

In this appendix we describe the effective theory, formed by taking the top quark mass to infinity~\cite{Shifman:1979eb,Voloshin:1985tc,Ellis:1975ap} which we use to compute the next-to-leading order corrections to central exclusive Higgs production. This approach has been found to give good agreement with the full theory, provided that the Higgs mass satisfies $m_H \lesssim 2 m_{\textrm{top}}$ and the transverse momenta of any jets produced in association with the Higgs satisfy $p_\perp \lesssim m_{\textrm{top}}$~\cite{DelDuca:2001fn,DelDuca:2003ba}.

We work in a theory in which the top quark has been integrated out and all other quarks are taken as massless. In this approach, the only coupling of the Higgs is to gluons\footnote{We consistently ignore electroweak couplings throughout.}, via the following term in the effective Lagrangian 
\begin{equation} \label{eq:LeffBare}
  \mathcal{L}_{\text{eff}}=-\frac{H}{4}C^0_1\mathcal{O}^0_1\;, \qquad \mathcal{O}^0_1=(G_0)^a_{\mu\nu}(G_0)^{a \mu\nu}
\end{equation}
where $C^0_1$ is a coefficient function, zeroes indicate that these are bare quantities and it is understood that they are defined in the five flavour effective theory. Both $C^0_1$ and matrix elements of $\mathcal{O}^0_1$ contain ultraviolet divergences, however their product is finite,
%(I have seen this stated and there must be some simple way of seeing this is the case, but at the moment it is eluding me!).
since the operator in the full theory which (\ref{eq:LeffBare}) approximates ($\frac{H}{v}m_t\bar{\psi}\psi$) is a conserved current. 

Using the Bogolyubov-Parasiuk R-operation~\cite{Bogolyubov:1980nc,Spiridonov:1984br}, it is possible to define a finite version of $\mathcal{O}^0_1$, which we denote $\mathcal{O}^R_1$. This finite operator may then be related to the bare operator as~\cite{Spiridonov:1984br}
\begin{equation}
	\mathcal{O}^R_1 = Z_{\mathcal{O}_1} \mathcal{O}^0_1\;, \qquad
	Z_{\mathcal{O}_1} =  \frac{1}{1-\beta(\mathcal{N}\alpha_s)/\epsilon}
\end{equation}
where 
\begin{equation}
	\mathcal{N} = \text{exp}[\epsilon(-\gamma_{\text{E}} + \ln(4\pi))],
\end{equation}
and $\beta(\alpha_s)$ and $\alpha_s \equiv \alpha_s(\mu)$ are the QCD beta function and the $\overline{\textrm{MS}}$ running coupling respectively. Again, both are defined in the five flavour theory.

The effective Lagrangian now reads

\begin{equation}
	\mathcal{L}_{\text{eff}}=-\frac{H}{4}C^R_1\mathcal{O}^R_1\; \qquad C^R_1 = Z_{\mathcal{O}_1}^{-1} C^0_1 \label{eq:RewrittenLeff}\;.
\end{equation}
Since both matrix elements of $\mathcal{O}^R_1$ and the full expression are finite, $C^R_1$ is also finite and is given by
\begin{equation}
	C^R_1(\mu) = -\frac{1}{3v}\frac{\alpha_s(\mu)}{\pi}\left( 1 + \frac{11}{4}\frac{\alpha_s(\mu)}{\pi}   \right) + \mathcal{O}(\alpha_s^3)
\end{equation}
where again $\alpha_s(\mu)$ is the five flavour $\overline{\textrm{MS}}$ running coupling and $v$ is the Higgs vacuum expectation value.

%%%%%%%%%%%%%%%%%%%%%%%%%%%%%%%%%%%%%%%%%
\begin{figure}[t]
	 \subfigure{\includegraphics[width=0.8\textwidth]{FeynmanDiagrams/LargeTopFeynman/1/1.epsi} }\\
	 \subfigure{\includegraphics[width=0.7\textwidth]{FeynmanDiagrams/LargeTopFeynman/2/2.epsi}}\\
	\subfigure{\includegraphics[width=0.7\textwidth]{FeynmanDiagrams/LargeTopFeynman/3/3.epsi}}
	\caption{Feynman rules for the large top mass effective theory. See the text for the definitions of $V_3$ and $V_4$.}\label{fig:LargeTopFeynman}
\end{figure}
%%%%%%%%%%%%%%%%%%%%%%%%%%%%%%%%%%%%%%%%%

The Feynman rules generated by equation~(\ref{eq:RewrittenLeff}) are displayed in figure~\ref{fig:LargeTopFeynman}, and can be written in terms of the standard three and four gluon vertices:
\begin{align}
	V_{3,\mu_1\mu_2\mu_3}^{a_1a_2a_3}(k_1,k_2,k_3) &= g f^{a_1a_2a_3}\big( g_{\mu_1\mu_2}(k_1-k_2)_{\mu_3} + g_{\mu_2\mu_3}(k_2-k_3)_{\mu_1} \nonumber \\
	& \qquad \qquad \qquad + g_{\mu_3\mu_1}(k_3-k_1)_{\mu_2} \big) \\
	V_{4,\mu_1\mu_2\mu_3\mu_4}^{a_1a_2a_3a_4}(k_1,k_2,k_3,k_4) &= -i g^2 \big( f^{a_1a_2e}f^{a_4a_3e}( g_{\mu_1\mu_4}g_{\mu_2\mu_3} - g_{\mu_1\mu_3}g_{\mu_4\mu_2}  ) \nonumber \\
	&  \qquad \qquad  f^{a_1a_3e}f^{a_4a_2e}( g_{\mu_1\mu_4}g_{\mu_2\mu_3} - g_{\mu_1\mu_2}g_{\mu_4\mu_3}  ) \nonumber \\
	&  \qquad \qquad  f^{a_1a_4e}f^{a_2a_3e}( g_{\mu_1\mu_2}g_{\mu_4\mu_3} - g_{\mu_1\mu_3}g_{\mu_4\mu_2}  )  \big) \;.
\end{align}

%% file: paper.bbl
\providecommand{\href}[2]{#2}\begingroup\raggedright\begin{thebibliography}{10}

\bibitem{Khoze:2000jm}
V.~A. Khoze, A.~D. Martin, and M.~G. Ryskin, {\it {Double-diffractive processes
  in high-resolution missing- mass experiments at the Tevatron}},  {\em Eur.
  Phys. J.} {\bf C19} (2001) 477--483,
  [\href{http://xxx.lanl.gov/abs/hep-ph/0011393}{{\tt hep-ph/0011393}}].

\bibitem{Albrow:2000na}
M.~G. Albrow and A.~Rostovtsev, {\it {Searching for the Higgs at hadron
  colliders using the missing mass method}},
  \href{http://xxx.lanl.gov/abs/hep-ph/0009336}{{\tt hep-ph/0009336}}.

\bibitem{Albrow:2008pn}
{\bf FP420 R \& D} Collaboration, M.~G. Albrow {\em et.~al.}, {\it {The FP420
  R\&D Project: Higgs and New Physics with forward protons at the LHC}},  {\em
  JINST} {\bf 4} (2009) T10001, [\href{http://xxx.lanl.gov/abs/0806.0302}{{\tt
  arXiv:0806.0302}}].

\bibitem{Kaidalov:2003fw}
A.~B. Kaidalov, V.~A. Khoze, A.~D. Martin, and M.~G. Ryskin, {\it {Central
  exclusive diffractive production as a spin parity analyser: From hadrons to
  Higgs}},  {\em Eur. Phys. J.} {\bf C31} (2003) 387--396,
  [\href{http://xxx.lanl.gov/abs/hep-ph/0307064}{{\tt hep-ph/0307064}}].

\bibitem{:2007na}
{\bf CDF} Collaboration, T.~Aaltonen {\em et.~al.}, {\it {Search for exclusive
  $\gamma \gamma$ production in hadron- hadron collisions}},  {\em Phys. Rev.
  Lett.} {\bf 99} (2007) 242002, [\href{http://xxx.lanl.gov/abs/0707.2374}{{\tt
  arXiv:0707.2374}}].

\bibitem{Aaltonen:2007hs}
{\bf CDF} Collaboration, T.~Aaltonen {\em et.~al.}, {\it {Observation of
  Exclusive Dijet Production at the Fermilab Tevatron $p\bar{p}$ Collider}},
  {\em Phys. Rev.} {\bf D77} (2008) 052004,
  [\href{http://xxx.lanl.gov/abs/0712.0604}{{\tt arXiv:0712.0604}}].

\bibitem{Aaltonen:2009kg}
{\bf CDF} Collaboration, T.~Aaltonen {\em et.~al.}, {\it {Observation of
  exclusive charmonium production and $\gamma\gamma$ to $\mu^+$$\mu^-$ in
  $p\bar{p}$ collisions at $\sqrt{s} = 1.96~\textrm{TeV}$}},  {\em Phys. Rev.
  Lett.} {\bf 102} (2009) 242001,
  [\href{http://xxx.lanl.gov/abs/0902.1271}{{\tt arXiv:0902.1271}}].

\bibitem{Khoze:2000cy}
V.~A. Khoze, A.~D. Martin, and M.~G. Ryskin, {\it {Can the Higgs be seen in
  rapidity gap events at the Tevatron or the LHC?}},  {\em Eur. Phys. J.} {\bf
  C14} (2000) 525--534, [\href{http://xxx.lanl.gov/abs/hep-ph/0002072}{{\tt
  hep-ph/0002072}}].

\bibitem{Khoze:2001xm}
V.~A. Khoze, A.~D. Martin, and M.~G. Ryskin, {\it {Prospects for new physics
  observations in diffractive processes at the LHC and Tevatron}},  {\em Eur.
  Phys. J.} {\bf C23} (2002) 311--327,
  [\href{http://xxx.lanl.gov/abs/hep-ph/0111078}{{\tt hep-ph/0111078}}].

\bibitem{Bjorken:1992er}
J.~D. Bjorken, {\it {Rapidity gaps and jets as a new physics signature in very
  high-energy hadron hadron collisions}},  {\em Phys. Rev.} {\bf D47} (1993)
  101--113.

\bibitem{Bartels:2006ea}
J.~Bartels, S.~Bondarenko, K.~Kutak, and L.~Motyka, {\it {Exclusive Higgs boson
  production at the LHC: Hard rescattering corrections}},  {\em Phys. Rev.}
  {\bf D73} (2006) 093004, [\href{http://xxx.lanl.gov/abs/hep-ph/0601128}{{\tt
  hep-ph/0601128}}].

\bibitem{Khoze:2006uj}
V.~A. Khoze, A.~D. Martin, and M.~G. Ryskin, {\it {On the role of hard
  rescattering in exclusive diffractive Higgs production}},  {\em JHEP} {\bf
  05} (2006) 036, [\href{http://xxx.lanl.gov/abs/hep-ph/0602247}{{\tt
  hep-ph/0602247}}].

\bibitem{Strikman:2008er}
M.~Strikman and C.~Weiss, {\it {Rapidity gap survival in central exclusive
  diffraction: Dynamical mechanisms and uncertainties}},
  \href{http://xxx.lanl.gov/abs/0812.1053}{{\tt arXiv:0812.1053}}.

\bibitem{Gotsman:2008tr}
E.~Gotsman, E.~Levin, U.~Maor, and J.~S. Miller, {\it {A QCD motivated model
  for soft interactions at high energies}},  {\em Eur. Phys. J.} {\bf C57}
  (2008) 689--709, [\href{http://xxx.lanl.gov/abs/0805.2799}{{\tt
  arXiv:0805.2799}}].

\bibitem{Gotsman:2009bn}
E.~Gotsman, E.~Levin, U.~Maor, and J.~S. Miller, {\it {Soft interactions at
  high energies: QCD motivated approach}},
  \href{http://xxx.lanl.gov/abs/0901.1540}{{\tt arXiv:0901.1540}}.

\bibitem{Ryskin:2009tk}
M.~G. Ryskin, A.~D. Martin, and V.~A. Khoze, {\it {Soft processes at the LHC,
  II: Soft-hard factorization breaking and gap survival}},  {\em Eur. Phys. J.}
  {\bf C60} (2009) 265--272, [\href{http://xxx.lanl.gov/abs/0812.2413}{{\tt
  arXiv:0812.2413}}].

\bibitem{Martin:2001ms}
A.~D. Martin and M.~G. Ryskin, {\it {Unintegrated generalised parton
  distributions}},  {\em Phys. Rev.} {\bf D64} (2001) 094017,
  [\href{http://xxx.lanl.gov/abs/hep-ph/0107149}{{\tt hep-ph/0107149}}].

\bibitem{Belitsky:2005qn}
A.~V. Belitsky and A.~V. Radyushkin, {\it {Unraveling hadron structure with
  generalized parton distributions}},  {\em Phys. Rept.} {\bf 418} (2005)
  1--387, [\href{http://xxx.lanl.gov/abs/hep-ph/0504030}{{\tt
  hep-ph/0504030}}].

\bibitem{Shuvaev:1999ce}
A.~G. Shuvaev, K.~J. Golec-Biernat, A.~D. Martin, and M.~G. Ryskin, {\it
  {Off-diagonal distributions fixed by diagonal partons at small $x$ and
  $\xi$}},  {\em Phys. Rev.} {\bf D60} (1999) 014015,
  [\href{http://xxx.lanl.gov/abs/hep-ph/9902410}{{\tt hep-ph/9902410}}].

\bibitem{Dokshitzer:1978hw}
Y.~L. Dokshitzer, D.~Diakonov, and S.~I. Troian, {\it {Hard Processes in
  Quantum Chromodynamics}},  {\em Phys. Rept.} {\bf 58} (1980) 269--395.

\bibitem{Fadin:1975cb}
V.~S. Fadin, E.~A. Kuraev, and L.~N. Lipatov, {\it {On the Pomeranchuk
  Singularity in Asymptotically Free Theories}},  {\em Phys. Lett.} {\bf B60}
  (1975) 50--52.

\bibitem{Kuraev:1976ge}
E.~A. Kuraev, L.~N. Lipatov, and V.~S. Fadin, {\it {Multi - Reggeon Processes
  in the Yang-Mills Theory}},  {\em Sov. Phys. JETP} {\bf 44} (1976) 443--450.

\bibitem{Kuraev:1977fs}
E.~A. Kuraev, L.~N. Lipatov, and V.~S. Fadin, {\it {The Pomeranchuk Singularity
  in Nonabelian Gauge Theories}},  {\em Sov. Phys. JETP} {\bf 45} (1977)
  199--204.

\bibitem{Balitsky:1978ic}
I.~I. Balitsky and L.~N. Lipatov, {\it {The Pomeranchuk Singularity in Quantum
  Chromodynamics}},  {\em Sov. J. Nucl. Phys.} {\bf 28} (1978) 822--829.

\bibitem{Forshaw:1997dc}
J.~R. Forshaw and D.~A. Ross, {\it {Quantum {C}hromodynamics and the
  {P}omeron}},  {\em Cambridge Lect. Notes Phys.} {\bf 9} (1997) 1--248.

\bibitem{Khoze:2008cx}
V.~A. Khoze, A.~D. Martin, and M.~G. Ryskin, {\it {Early LHC measurements to
  check predictions for central exclusive production}},  {\em Eur. Phys. J.}
  {\bf C55} (2008) 363--375, [\href{http://xxx.lanl.gov/abs/0802.0177}{{\tt
  arXiv:0802.0177}}].

\bibitem{Kaidalov:2003ys}
A.~B. Kaidalov, V.~A. Khoze, A.~D. Martin, and M.~G. Ryskin, {\it {Extending
  the study of the Higgs sector at the LHC by proton tagging}},  {\em Eur.
  Phys. J.} {\bf C33} (2004) 261--271,
  [\href{http://xxx.lanl.gov/abs/hep-ph/0311023}{{\tt hep-ph/0311023}}].

\bibitem{Shifman:1979eb}
M.~A. Shifman, A.~I. Vainshtein, M.~B. Voloshin, and V.~I. Zakharov, {\it
  {Low-Energy Theorems for Higgs Boson Couplings to Photons}},  {\em Sov. J.
  Nucl. Phys.} {\bf 30} (1979) 711--716.

\bibitem{Voloshin:1985tc}
M.~B. Voloshin, {\it {Once Again About the Role of Gluonic Mechanism in
  Interaction of Light Higgs Boson with Hadrons}},  {\em Sov. J. Nucl. Phys.}
  {\bf 44} (1986) 478.

\bibitem{Ellis:1975ap}
J.~R. Ellis, M.~K. Gaillard, and D.~V. Nanopoulos, {\it {A Phenomenological
  Profile of the Higgs Boson}},  {\em Nucl. Phys.} {\bf B106} (1976) 292.

\bibitem{Bodwin:1981fv}
G.~T. Bodwin, S.~J. Brodsky, and G.~P. Lepage, {\it {Initial State Interactions
  and the Drell-Yan Process}},  {\em Phys. Rev. Lett.} {\bf 47} (1981) 1799.

\bibitem{Collins:1981ta}
J.~C. Collins and G.~Sterman, {\it {Soft partons in QCD}},  {\em Nucl. Phys.}
  {\bf B185} (1981) 172.

\bibitem{Cutkosky:1960sp}
R.~E. Cutkosky, {\it {Singularities and discontinuities of Feynman
  amplitudes}},  {\em J. Math. Phys.} {\bf 1} (1960) 429--433.

\bibitem{Collins:1985ue}
J.~C. Collins, D.~E. Soper, and G.~Sterman, {\it {Factorization for Short
  Distance Hadron-Hadron Scattering}},  {\em Nucl. Phys.} {\bf B261} (1985)
  104.

\bibitem{Collins:1988ig}
J.~C. Collins, D.~E. Soper, and G.~Sterman, {\it {Soft Gluons and
  Factorization}},  {\em Nucl. Phys.} {\bf B308} (1988) 833.

\bibitem{Collins:1981uk}
J.~C. Collins and D.~E. Soper, {\it {Back-To-Back Jets in QCD}},  {\em Nucl.
  Phys.} {\bf B193} (1981) 381.

\bibitem{Collins:1989gx}
J.~C. Collins, D.~E. Soper, and G.~Sterman, {\it {Factorization of Hard
  Processes in QCD}},  {\em Adv. Ser. Direct. High Energy Phys.} {\bf 5} (1988)
  1--91, [\href{http://xxx.lanl.gov/abs/hep-ph/0409313}{{\tt hep-ph/0409313}}].

\bibitem{Cudell:2008gv}
J.~R. Cudell, A.~Dechambre, O.~F. Hernandez, and I.~P. Ivanov, {\it {Central
  exclusive production of dijets at hadronic colliders}},  {\em Eur. Phys. J.}
  {\bf C61} (2009) 369--390, [\href{http://xxx.lanl.gov/abs/0807.0600}{{\tt
  arXiv:0807.0600}}].

\bibitem{Gunion:1976iy}
J.~F. Gunion and D.~E. Soper, {\it {Quark Counting and Hadron Size Effects for
  Total Cross- Sections}},  {\em Phys. Rev.} {\bf D15} (1977) 2617--2621.

\bibitem{Levin:1981rf}
E.~M. Levin and M.~G. Ryskin, {\it {Born Approximation of {QCD} for Description
  of High-Energy Hadronic Interactions}},  {\em Sov. J. Nucl. Phys.} {\bf 34}
  (1981) 619--623.

\bibitem{Catani:1990rr}
S.~Catani, B.~R. Webber, and G.~Marchesini, {\it {QCD coherent branching and
  semiinclusive processes at large $x$}},  {\em Nucl. Phys.} {\bf B349} (1991)
  635--654.

\bibitem{Grozin:2007zz}
A.~Grozin, {\it {Lectures on QED and QCD: Practical calculation and
  renormalization of one- and multi-loop Feynman diagrams}}, . Hackensack, USA:
  World Scientific (2007) 224 p.

\bibitem{Bassetto:1984ik}
A.~Bassetto, M.~Ciafaloni, and G.~Marchesini, {\it {Jet Structure and Infrared
  Sensitive Quantities in Perturbative QCD}},  {\em Phys. Rept.} {\bf 100}
  (1983) 201--272.

\bibitem{Spira:1995rr}
M.~Spira, A.~Djouadi, D.~Graudenz, and P.~M. Zerwas, {\it {Higgs boson
  production at the LHC}},  {\em Nucl. Phys.} {\bf B453} (1995) 17--82,
  [\href{http://xxx.lanl.gov/abs/hep-ph/9504378}{{\tt hep-ph/9504378}}].

\bibitem{Kunszt:1996yp}
Z.~Kunszt, S.~Moretti, and W.~J. Stirling, {\it {Higgs production at the LHC:
  An update on cross sections and branching ratios}},  {\em Z. Phys.} {\bf C74}
  (1997) 479--491, [\href{http://xxx.lanl.gov/abs/hep-ph/9611397}{{\tt
  hep-ph/9611397}}].

\bibitem{Ellis:2005zh}
R.~K. Ellis, W.~T. Giele, and G.~Zanderighi, {\it {Semi-numerical evaluation of
  one-loop corrections}},  {\em Phys. Rev.} {\bf D73} (2006) 014027,
  [\href{http://xxx.lanl.gov/abs/hep-ph/0508308}{{\tt hep-ph/0508308}}].

\bibitem{Mathematica}
{Wolfram Research Inc.}, {\it {Mathematica, Version 6.0}}, .

\bibitem{FORM}
J.~Vermaseren, {\it {New features of FORM}},
  \href{http://xxx.lanl.gov/abs/math-ph/0010025}{{\tt math-ph/0010025}}.

\bibitem{Mertig:1990an}
R.~Mertig, M.~Bohm, and A.~Denner, {\it {FEYN CALC: Computer algebraic
  calculation of Feynman amplitudes}},  {\em Comput. Phys. Commun.} {\bf 64}
  (1991) 345--359.

\bibitem{Bloch:1937pw}
F.~Bloch and A.~Nordsieck, {\it {Note on the Radiation Field of the electron}},
   {\em Phys. Rev.} {\bf 52} (1937) 54--59.

\bibitem{Cutler:1977qm}
R.~Cutler and D.~W. Sivers, {\it {Quantum Chromodynamic Gluon Contributions to
  Large p(T) Reactions}},  {\em Phys. Rev.} {\bf D17} (1978) 196.

\bibitem{Monk:2005ji}
J.~Monk and A.~Pilkington, {\it {ExHuME: A Monte Carlo event generator for
  exclusive diffraction}},  {\em Comput. Phys. Commun.} {\bf 175} (2006)
  232--239, [\href{http://xxx.lanl.gov/abs/hep-ph/0502077}{{\tt
  hep-ph/0502077}}].

\bibitem{MRST}
A.~D. Martin, R.~G. Roberts, W.~J. Stirling, and R.~S. Thorne, {\it
  Uncertainties of predictions from parton distributions {I}: Experimental
  errors},  {\em Eur. Phys. J.} {\bf C28} (2003) 455--473,
  [\href{http://xxx.lanl.gov/abs/hep-ph/0211080}{{\tt hep-ph/0211080}}].

\bibitem{CTEQ}
J.~Pumplin {\em et.~al.}, {\it New generation of parton distributions with
  uncertainties from global {QCD} analysis},  {\em JHEP} {\bf 07} (2002) 012,
  [\href{http://xxx.lanl.gov/abs/hep-ph/0201195}{{\tt hep-ph/0201195}}].

\bibitem{Forshaw:2005qp}
J.~R. Forshaw, {\it Diffractive {H}iggs production: {T}heory},
  \href{http://xxx.lanl.gov/abs/hep-ph/0508274}{{\tt hep-ph/0508274}}.

\bibitem{DelDuca:2001fn}
V.~Del~Duca, W.~Kilgore, C.~Oleari, C.~Schmidt, and D.~Zeppenfeld, {\it
  {Gluon-fusion contributions to H + 2 jet production}},  {\em Nucl. Phys.}
  {\bf B616} (2001) 367--399,
  [\href{http://xxx.lanl.gov/abs/hep-ph/0108030}{{\tt hep-ph/0108030}}].

\bibitem{DelDuca:2003ba}
V.~Del~Duca, W.~Kilgore, C.~Oleari, C.~R. Schmidt, and D.~Zeppenfeld, {\it
  {Kinematical limits on Higgs boson production via gluon fusion in association
  with jets}},  {\em Phys. Rev.} {\bf D67} (2003) 073003,
  [\href{http://xxx.lanl.gov/abs/hep-ph/0301013}{{\tt hep-ph/0301013}}].

\bibitem{Bogolyubov:1980nc}
N.~N. Bogolyubov and D.~V. Shirkov, {\it Introduction to the theory of
  quantized fields},  {\em Intersci. Monogr. Phys. Astron.} {\bf 3} (1959)
  1--720.

\bibitem{Spiridonov:1984br}
V.~P. Spiridonov, {\it Anomalous dimension of {$G_{\mu\nu}^2$} and
  $\beta$-function}, . IYaI-P-0378.

\end{thebibliography}\endgroup
